\documentclass[prb,twocolumn,amsmath,amssymb,floatfix,eqsecnum]{revtex4}
\usepackage{graphicx}
\usepackage{bm}

\begin{document}

\title{Non-equilibrium Gross-Pitaevskii dynamics of boson lattice models}

\author{Anatoli Polkovnikov}
\email{anatoli.polkovnikov@yale.edu}
\homepage{http://pantheon.yale.edu/~asp28} \affiliation{Department
of Physics, Yale University, P.O. Box 208120, New Haven CT
06520-8120}

\author{Subir Sachdev}
\email{subir.sachdev@yale.edu}
\homepage{http://pantheon.yale.edu/~subir} \affiliation{Department
of Physics, Yale University, P.O. Box 208120, New Haven CT
06520-8120}

\author{S. M. Girvin}
\email{steven.girvin@yale.edu}
\homepage{http://pantheon.yale.edu/~smg47} \affiliation{Department
of Physics, Yale University, P.O. Box 208120, New Haven CT
06520-8120}

\date{\today}

\begin{abstract}
Motivated by recent experiments on trapped ultra-cold
bosonic atoms in an optical lattice potential, we consider the
non-equilibrium dynamic properties of such bosonic systems
for a number of experimentally
relevant situations. When the number of bosons per lattice
site is large, there is a wide parameter regime where
the effective boson interactions are strong, but the
ground state remains a superfluid (and not a Mott insulator):
we describe the conditions under which the dynamics in
this regime can be described by a discrete
Gross-Pitaevskii equation. We describe the evolution of the
phase coherence after the system is initially prepared
in a Mott insulating state, and then allowed to evolve after a
sudden change in parameters places it in a regime with a
superfluid ground state. We also consider initial conditions with
a ``$\pi$ phase'' imprint on a superfluid ground state ({\em i.e.} the
initial phases of neighboring wells differ by $\pi$), and
discuss the subsequent appearance of density wave order
and ``Schr\"odinger cat'' states.
\end{abstract}

\maketitle

\section{Introduction}
\label{sec:intro} With the emerging experimental studies of
ultra-cold atoms in a parabolic trap and a periodic optical
lattice potential \cite{Kasevich,Bloch} (the wavelength of the
optical potential is much smaller than the dimensions of the
trap), new possibilities for studying the physics of interacting
bosons have emerged. At equilibrium, the bosons can undergo a
transition from a superfluid to an insulator as the strength of
the optical potential is
increased~\cite{Fisher,Sachdev,Monien,Elstner,Amico,Jaksh}.
However, the facile tunability and long characteristic time scales
of these systems also offer an opportunity to investigate
non-equilibrium dynamical regimes that have not been accessible
before. In this context, there have been a few recent theoretical
studies of the dynamics of bosons in a periodic potential:
Ref.~\onlinecite{Stringari} computed the oscillation frequency of
the center of mass of a superfluid state of bosons, while some
non-equilibrium issues were addressed in papers
\cite{Franzosi,Bishop,Bishop2} which appeared while this paper was
being completed.

A description of the purpose of this paper requires an
understanding of the different parameter regimes of the boson
system, which we will assume is well described by the single-band
Hubbard model:
\begin{eqnarray}
\mathcal{H}&=&\sum_j \Biggl[ -J(a_j^\dagger
a_{j+1}+a_{j+1}^\dagger a_j )+V_j a_j^\dagger a_j\nonumber \\
&~&~~~~~~+ {U\over 2}
a_j^\dagger a_j (a_j^\dagger a_j-1) \Biggr]. \label{1}
\end{eqnarray}
Here $a_j$ is a canonical Bose annihilation operator on sites
of the optical lattice (``wells'') labeled by the integer $j$, $J$ is the
tunneling amplitude between neighboring lattice sites, $U>0$ in the
repulsive interaction energy between bosons in the same lattice
minimum, and $V_j$  is a smooth external potential which we will
take to be parabolic. We will mainly consider the case of a
one-dimensional optical lattice, relevant to the experiments of
Ref.~\onlinecite{Kasevich}, but generalization to higher dimensions is
possible. The form of $V_j$ and the chemical potential of the
bosons determine another important parameter: $N$, the mean number
of bosons at the central site (more precisely, at the site
where $V_j$ is smallest);
we shall mainly consider the case $N \gg 1$
here. A dimensionless measure of the strength of the interactions
between the bosons is the coupling
\begin{equation}
\lambda \equiv \frac{UN}{J}; \label{deflambda}
\end{equation}
the different physical regimes of $\mathcal{H}$ are also conveniently
dilineated by the values of $\lambda$.

When the interactions between the bosons are strong enough,
$\lambda > \lambda_{SI}$, the ground state of $\mathcal{H}$
undergoes a quantum phase transition from a superfluid to a Mott
insulator (see Appendix~\ref{app:mft}). It is known that
\cite{Fisher}: \begin{equation} \lambda_{SI} \sim N^2 .
\end{equation} So for the case where $N$ is large, there is a wide
regime, $1 \ll \lambda \ll N^2$, where the interactions between
the bosons are very strong, but the ground state is nevertheless a
superfluid. A description of the dynamical properties of
$\mathcal{H}$ in this regime is one of central purposes of this
paper.

For $N$ large, and $\lambda$ smaller than $\lambda_{SI}$, it is
widely accepted \cite{Franzosi} that the low temperature dynamics
of $\mathcal{H}$ can be described by treating the operator $a_{j}$
as a classical $c$-number. (We will investigate the conditions for
the validity of this classical approximation more carefully in
Section~\ref{sec:two}, where we will also discuss the time range
over which it can be applied.) More precisely, we introduce the
dimensionless complex dynamical variable $\psi_j (t)$ whose value
is a measure of $\langle a_j (t) \rangle/\sqrt{N}$; then its
dynamics is described by the classical Hamiltonian
\begin{equation} H_{GP}=\sum_j \left[ - (\psi_j^\star
\psi_{j+1}+\psi_{j+1}^\star\psi_j)+ \frac{V_j}{J}
|\psi_j|^2+{\lambda \over 2}|\psi_j|^4 \right],\label{1b}
\end{equation} and the Poisson brackets \begin{equation} \left\{ \psi_j,
\psi_j^{\star} \right\} = \delta_{ij}. \label{poisson}
\end{equation} Here, and henceforth, we measure time in units of
$\hbar/J$. The resulting equations of motion are, of course, a
discrete version of the familiar Gross-Pitaevskii (GP) equations.
We will often impose a parabolic confining potential, in which
case
\begin{displaymath}
{V_j\over J}={\xi\over 2} j^2 . \label{defxi}
\end{displaymath}
A nonuniform potential $V_j$ also can lead to localization of
bosons in separate wells; in particular, even without interaction
($\lambda=0$), when $|V_{j+1}-V_j|\geq 2J$ the eigenmodes of
($\ref{1b}$) become localized. Note that this localization is a
purely semiclassical effect, described by the GP equations. If
$V_j$ is smooth then for $\lambda > \lambda_{SI}$, the system
undergoes a transition to nonuniform insulating
state~\cite{Troyer,Svistunov}.

Describing the non-equilibrium quantum Bose dynamics for
$\lambda < \lambda_{SI}$ is now reduced
to a problem of integrating the classical equations of motion implied
by (\ref{1b},\ref{poisson}). However, it remains to specify the
initial conditions for the classical equations; these clearly depend
upon the physical situations of interest, and we shall consider here
two distinct cases, which are discussed in the following subsections

\subsection{Mott insulating initial state}
\label{sec:mottini}

Consider the physical situation (of current experimental
interest~\cite{Kasevich2}) where for  $t \leq 0$ the bosons are in
a Mott insulating state with $\lambda > \lambda_{SI}$, and at time
$t=0$ the optical lattice potential is suddenly reduced so that
$\lambda < \lambda_{SI}$ for all $t > 0$. Clearly, the GP
equations should apply for $t >0$, and the Mott insulating initial
state will impose initial conditions which we now describe. The
required initial conditions are readily deduced by thinking about
the full quantum Heisenberg equations of motion for $a_{j} (t)$
implied by $\mathcal{H}$. By integrating these equations, one can,
in principle, relate any observable to the expectation values of
products of powers of $a_{j}^{\dagger} (t=0)$ and $a_{j} (t=0)$.
For the Mott insulator with $\lambda \gg \lambda_{SI}$ these
expectation values have a very simple structure: they factorize
into products of expectation values on each site, and are non-zero
only if the number of creation and annihilation operators on each
site are equal. Furthermore, for large $N$, we can also ignore the
ordering of the $a_j$ and $a_j^{\dagger}$ operators on each site,
and {\em e.g.} we obtain to leading order in $1/N$:
\begin{equation} \left\langle a_{j}^{\dagger n} (t=0) a_{\ell}^m
(t=0) \right \rangle \approx \delta_{nm} \delta_{j\ell}
\left(N_j\right)^n, \label{ev} \end{equation} where we have
accounted for a possible spatial inhomogeneity by introducing
$N_j$ (a number of order $N$), the number of bosons at site $j$ in
the Mott insulator. In terms of the classical variables $\psi_j$,
the $t=0$ expectation values in (\ref{ev}) are easy to reproduce.
We simply choose \begin{equation} \psi_j (t=0) = \sqrt{N_j / N}
e^{i \phi_j} \label{ev1}
\end{equation} where the $\phi_j$ are {\em independent random
variables} which are uniformly distributed between $0$ and $2
\pi$. In this manner, we have mapped the fully deterministic
quantum time evolution of $\mathcal{H}$ to the stochastic and
classical time evolution of $H_{GP}$. In practice, the procedure
is then as follows: choose a large ensemble of initial values of
$\phi_j$, and deterministically evolve $H_{GP}$ for each such
initial condition; the expectation value of any quantum observable
at time $t$ is then given by the average value of the
corresponding classical observable at time $t$, with the average
being taken over the random variables $\phi_j$. In particular
\begin{equation} \left\langle a_{j}^{\dagger n} (t) a_{j'}^m (t)
\right \rangle_Q \approx N \left\langle \psi_j^{\star n}(t)
\psi_{j'}^m (t) \right \rangle_{\mbox{random $\phi_{\ell}$}},
\label{ev2} \end{equation} where we have indicated that the
angular brackets on the left represent a traditional quantum
expectation value, while those on the right represent an average
over the independent variables $\phi_j$ specified by (\ref{ev1})
at time $t=0$. We will henceforth implicitly assume that all
angular brackets have the meaning specified in (\ref{ev2}),
depending upon whether they contain quantum or classical
variables.

An important property of (\ref{ev2}) is that while we must have
$j\,' =j$ for a non-zero result at $t=0$, this is no longer true
for $t>0$. In particular, non-zero correlations can develop for
large $|j\, '-j|$ as time evolves, corresponding to a restoration
of phase coherence. Indeed the ground state for $\lambda <
\lambda_{SI}$ is superfluid and thermalization must lead to
increase of the phase correlations. However, in this paper we
show, that even without relaxation the coherence can be restored
dynamically. (Of course, as we are looking at one dimensional
systems and the final state is expected to be thermalized at a
non-zero temperature, the phase correlations cannot be truly
long-range and must decay exponentially at large enough scales:
however, guided by the experimental situation, we will look at
relatively small systems for which this is not an issue.)
Describing the dynamics of the restoration of this phase coherence
is also a central purpose of this paper. We shall characterize the
phase coherence by studying the expectation value of
\begin{equation}
\mathcal{D}_g (t) = \frac{1}{M} \sum_{j\neq \ell}
g (|j-\ell|) \left\langle \psi_j^{\star} (t) \psi_{\ell} (t) \right\rangle
\label{defDg}
\end{equation}
where $M$ is the number of lattice sites (for a nonuniform
external potential $V_j$, $M$ is just the ratio of the total
number of bosons to the number of bosons in central well), and
$g$ is some suitably chosen weight function. Observables closely
related to $\mathcal{D}_g$ are measured upon detecting the atoms
after releasing the trap. At time $t=0$, $\mathcal{D}_g (0) = 0$,
and we will be interested in the deviations of $\mathcal{D}_g (t)$
from this value for $t > 0$, an increase corresponding to an
enhancement of superfluid phase coherence. We note, in passing,
that a closely related procedure was used earlier \cite{Kedar} to
describe the onset of phase coherence after a sudden quench from
high temperature; here, we are always at zero temperature, and
move into a superfluid parameter regime by a sudden change in the
value of $\lambda$.

We will begin our analysis of the structure of $\mathcal{D}_g (t)$
by considering the case with two wells ($M=2$) in
Section~\ref{sec:twomott}. For the weakly interacting case
($\lambda \ll 1$), $\mathcal{D}_g (t)$ exhibits Josephson
oscillations with a period of order unity; the weak interactions
lead to a decay of oscillations with a slow ($t^{-1/2}$)
saturation of the coherence at a steady-state value at a time
scale $t\propto \lambda^{-1}$. For $\lambda\gg 1$ the oscillations
are suppressed and $\mathcal{D}_g (t)$ saturates at $t\propto
1/\sqrt\lambda $, which is, in fact, shorter than a single
tunneling time. For this two lattice site case we can also obtain
a complete solution for $\mathcal{D}_g (t)$ for the quantum
Hamiltonian $\mathcal{H}$ (described in
Section~\ref{sec:quantum}), and this allows a detailed analysis on
the regime of validity of the semiclassical GP equations. We show
that the semiclassical approach is valid for two lattice sites
when $N$ is large and $t<N/\lambda$. This is, in fact, a general
result which implies that the quantum mechanics becomes important
when time exceeds inverse energy level spacing. For more than two
lattice sites, the energy splitting scales as the inverse of the
total number of particles and at $\lambda\ll 1$, the semiclassical
conditions are virtually always fulfilled. It is surprising that
even with a small number of particles $N=4$, and weak
interactions, the GP equations give an excellent description of
the system evolution, apart from overall numerical prefactor
$(1+2/N)$, which is not small in this case.

The restoration of coherence is also studied in the many well
case in Section~\ref{sec:manymott}. We discuss the case without
an external potential in Section~\ref{sec:pbc}; with an equal
number of particles initially in all the wells, phase
correlations develop only in the interacting case ($\lambda>0$).
This is true  for both periodic and open boundary conditions.
Similar to the two well case, in the weakly interacting regime
phase correlations will oscillate in time. However these
oscillations will be periodic only for particular number of
wells: $M=2,3,4,6$ for periodic boundary conditions and $M=2,3,5$
for open boundary conditions. For other numbers of wells, the
oscillations are chaotic. As for the two well case, a stronger
interaction results in decay of correlations in time, leading to
the steady state.

Next, in Section~\ref{sec:parabola},
we consider the restoration of phase coherence
for the experimentally important case of a parabolic potential.
The results are quite different for this case, and
phase correlations develop even without interactions. In
a weak parabolic potential, $\mathcal{D}_g (t)$ oscillates
with a frequency which
scales as the square root of the parabolicity, $\xi$.
This frequency is closely related to the oscillation frequency
discussed recently by Kramer {\em et al.} \cite{Stringari} for the
case where the center of mass of the atomic gas is displaced.
In the present situation, there is no displacement
of the center of mass, but the same
oscillation is excited upon a sudden change in the value of
$\lambda$. The oscillations decay
even at $\lambda=0$; weak or intermediate interactions
$\lambda\leq 1$ do not change the noninteracting picture much.
The amplitude of the oscillations become more pronounced for
$\lambda\approx 1$, but for $\lambda\gg 1$
the oscillations are suppressed as for the flat potential.

While this work was being completed, we became aware of related
results of Altman and Auerbach also addressing the restoration of
phase coherence in a Mott insulator. However, there are some
significant differences in the physical situations being
addressed. Above, we have considered a system deep in the Mott
insulating phase (with $\lambda \gg \lambda_{SI}$) taken suddenly
to parameters for which the ground state was deep in the
superfluid phase (with $\lambda \ll \lambda_{SI}$). In contrast,
Ref.~\onlinecite{aa} consider the case when both the initial and
final values of $\lambda$ were not too far from $\lambda_{SI}$,
but remained on opposite sides of it. For $\lambda$ close to
$\lambda_{SI}$, and at temperatures not too small, a
``relativistic Gross-Pitaevski'' equation had been proposed in
Ref.~\onlinecite{ssrelax} as a description of the ``{\em Bose
molasses}'' dynamics of the order parameter. The conditions under
which oscillations in the amplitude of the order parameter would
be underdamped were also presented\cite{ssrelax}. Altman and
Auerbach \cite{aa} advocated that the same equations could
describe the time evolution of the amplitude of the order
parameter as it evolved from the Mott insulator (with zero
amplitude) to the superfluid (with finite amplitude) at zero
temperature. We review issues related to the damping of the
amplitude mode in Appendix~\ref{app:amp}. Altman and
Auerbach~\cite{aa} also considered the situation without an
external potential ($V_j\equiv 0$). We have noted above that such
a potential changed our results significantly; in
Appendix~\ref{app:mft} we discuss the significant role of the
external potential in the equilibrium properties for
$\lambda\approx \lambda_{SI}$.

\subsection{Modulated phase initial state}
\label{sec:modini}

A second set of initial conditions we consider is the case in
which the parameter values always correspond to a superfluid
ground state {\em i.e.} $\lambda < \lambda_{SI}$. For time $t \leq
0$ we imagine that $\lambda$ takes some fixed value and the phases
$\phi_j$ have some known set of fixed, {\em non-random} values at
$t=0$ and we follow the subsequent evolution of the bosons using
the discrete GP equation. The phase imprint can be experimentally
achieved by e.g. applying a short (compared to a single tunneling
time) pulse of external field to the condensate. A case of
special interest will be when there is a relative $\pi$ phase
shift between neighboring wells:
\begin{equation}
\phi_j = j \pi. \label{pi}
\end{equation}
For two wells with equal $N_j$ and relatively small $\lambda$,
this state is metastable (this is also the case for even $M$ and
periodic boundary conditions). However, if the interaction
$\lambda$ becomes larger than a critical value, this equilibrium
becomes unstable and the bosons spontaneously form a ``dipole''
state~\cite{Franzosi,Vardi,Raghavan} in which most of them occupy
one of the two wells (see Section~\ref{sec:twomod}). Upon
accounting for quantum tunneling in a system with a finite number
of bosons, the state obtained is a superposition of the two
dipole states restoring translational symmetry. However, in case
of infinite number of wells (see Section~\ref{sec:manymod}) the
tunneling between the two dipole configurations is negligible and
translational symmetry is broken by the appearance of a density
wave of bosons with a period of two lattice spacings. This effect
is similar to that studied in Ref.~\onlinecite{Sachdev1} for the
case of a Mott insulator in a strong electric field.

Related to this instability is a very interesting possibility of
forming a Schr\"odinger cat state~\cite{Kasevich1}. We show in
Section~\ref{sec:manymod} that if the system is initially in the
``$\pi$ state'', and the interaction is slowly increased, then at
certain point {\em all} the bosons spontaneously move into one of
the wells. If quantum mechanical corrections are taken into
account then the final configuration is the superposition of the
states with all bosons in one of the wells. This effect opens the
possibility of dynamical forming of a strongly entangled state of
bosons.

\section{Semiclassical versus quantum dynamics of two
coupled interacting Bose systems}
\label{sec:two}

The comparison between the semiclassical and quantum theory of the
two-well system has been presented earlier by Milburn {\em et al.}
\cite{milburn}, although for initial conditions different from
those we shall consider here.

First we will focus on the semiclassical description of the two
well system, when the total number of bosons is much greater than
1. In this case the Gross-Pitaevskii equations implied by
(\ref{1b}) and (\ref{poisson}) are
\begin{eqnarray}
&&i{\partial \psi_1\over \partial t}=-\psi_2+\lambda
|\psi_1|^2\psi_1,\label{2}\\
&&i{\partial \psi_2\over \partial t}=-\psi_1+\lambda
|\psi_2|^2\psi_2,\label{3}
\end{eqnarray}
The total number of bosons $|\psi_1|^2 + |\psi_2|^2$ is a
constant of the motion; with our normalization for $\psi_j$
described above (\ref{1b}), we have
$|\psi_1|^2+|\psi_2|^2=2$.

We use the parameterization: \begin{equation}
\psi_{1,2}=\sqrt{1\mp n}{\rm e^{i\theta\mp i\phi/2}}. \label{4}
\end{equation} Note that only the relative phase of $\psi_1$ and
$\psi_2$ is an observable. Substituting (\ref{4}) into (\ref{2})
and (\ref{3}) we obtain:
\begin{eqnarray}
&&{d^2 n\over dt^2} +4n+4\lambda n\sqrt{1-n^2}\cos\phi=0,\\
&&{d\cos\phi\over d n}={n\over 1-n^2}\cos\phi+{\lambda
n\over {\sqrt{1-n^2}}}.
\end{eqnarray}

After further manipulation this system reduces to a single second
order differential equation for the continuous variable $n$:

\begin{equation}
{d^2n\over dt^2}+4 n +4\lambda n \left(\cos\phi_0+{\lambda
n^2\over 2 }\right)=0\label{9}
\end{equation}
with initial conditions: $n(0)=n_0$, $dn(0)/dt=2\sin{\phi_0}$.
Similar equations were derived in~\cite{Franzosi,Raghavan}.
Without interaction ($\lambda=0$) we have a situation of a single
Josephson junction described by a free harmonic oscillator. The
interaction $\lambda$ is responsible for the anharmonicity. Note
that for $\lambda\leq 1$ the solutions $n=0$, $\phi=0,\pi$ are
stationary; {\em i.e.} the phase difference between the two wells
can be either $0$ or $\pi$. On the other hand for $\lambda>1$ the
solution with $\phi=\pi$ becomes
unstable~\cite{Franzosi,Raghavan}, and instead the new minima appear
at
\begin{equation}
n_{\rm min}=\pm \sqrt{2 (\lambda-1)\over \lambda^2}.
\end{equation}

We will now consider the properties of the two well system for the two
classes of initial conditions discussed in Section~\ref{sec:intro}
in turn. Each subsection below also contains a comparison
with the exact results obtained by a full quantum solution
of $\mathcal{H}$.

\subsection{Mott insulating initial state}
\label{sec:twomott}

As in Section~\ref{sec:mottini},
let us assume that initially the two condensates are completely
uncoupled. We will consider their evolution in the semiclassical and
quantum calculations in turn:

\subsubsection{Semiclassical theory}
\label{sec:semiclassical}

>From the discussion in Section~\ref{sec:mottini}, we have $n_0=0$
and $\phi_0$ is a uniform random variable. We will study the
correlation between $\psi_1$ and $\psi_2$ as a function of time.
It is easy to show that \begin{equation} \langle \psi_2^\star (t)
\psi_1(t)+\psi_1^\star(t)\psi_2(t)\rangle={\lambda\over 4}\langle
n^2(t)\rangle,\label{11} \end{equation} where the average is taken
over all possible initial phases $\phi_0$. The correlator is
proportional to the product of the coupling constant $\lambda$ and
the variance of $n$, reflecting the usual phase-number uncertainty
relation.

Before proceeding with quantitative analysis let us argue
qualitatively what happens with the system. Suppose $\lambda\ll
1$. Then (\ref{9}) is equivalent to the motion of a particle in a
harmonic potential with random initial velocity. Because the
frequency of the harmonic oscillator doesn't depend on the
amplitude, $\langle n^2(t)\rangle$ is a periodic function of time
with $T=\pi/2$. If $\lambda$ is still small but not negligible,
then (\ref{9}) still describes motion in a harmonic potential,
which, however, depends on the initial conditions. As a result the
oscillations of $\langle n^2(t)\rangle$ become quasiperiodic and
decay with time. In the limit of large $\lambda$ the oscillations
completely disappear and the steady state solution develops
during the time $t\sim 1/ \sqrt{\lambda}$.

For weak coupling $\lambda$, equation (\ref{9}) can be solved
explicitly. Thus for $\lambda=0$ \begin{equation} \langle
n^2(t)\rangle={1-\cos 4 t\over 4}.\label{12} \end{equation}

For small $\lambda$ the approximate analytical solution is:

\begin{equation}
\langle n^2(t)\rangle \approx {1\over 4}-{1\over 2\pi}
\int\limits_0^\pi \sin^2 \phi_0 \cos \left(4t\sqrt{1+\lambda\cos \phi_0 }\right)d\phi_0.
\end{equation}

It is easy to see that at large $t$ we have the following
asymptotic behavior:
\begin{eqnarray}
&&\langle n^2(t)\rangle \approx {1\over 4}- {1\over
\sqrt{16\pi\lambda t}} \left[\cos\left(4 t \sqrt{1+\lambda}
+ {\pi\over 4}\right)\right.\nonumber\\
&&\left. +\cos\left(4 t \sqrt{1-\lambda} - {\pi\over
4}\right)\right],\label{13} \end{eqnarray}
\begin{figure}
\includegraphics[width=7.5cm]{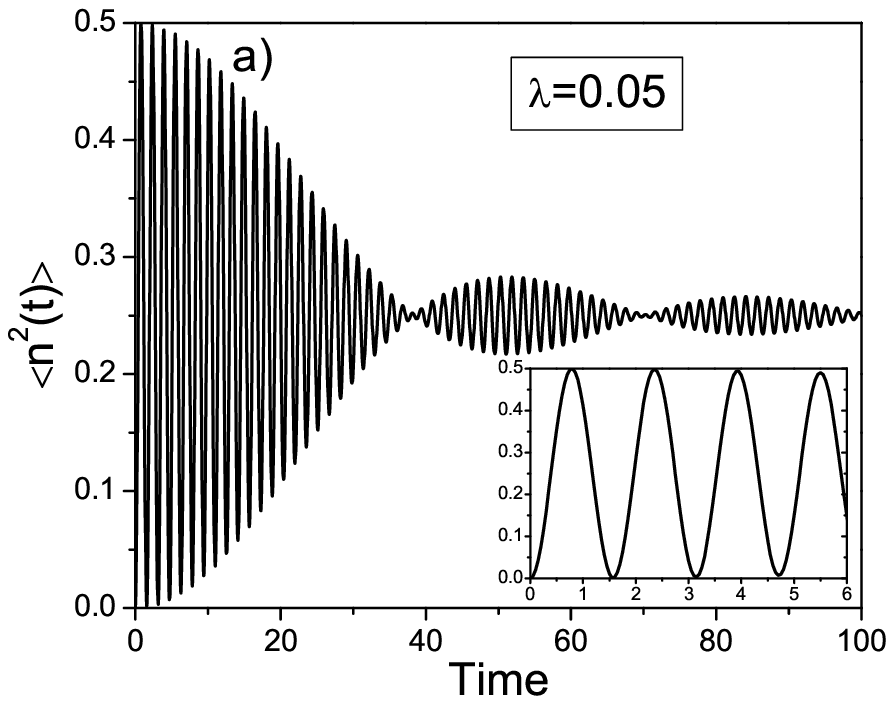}
\includegraphics[width=7.5cm]{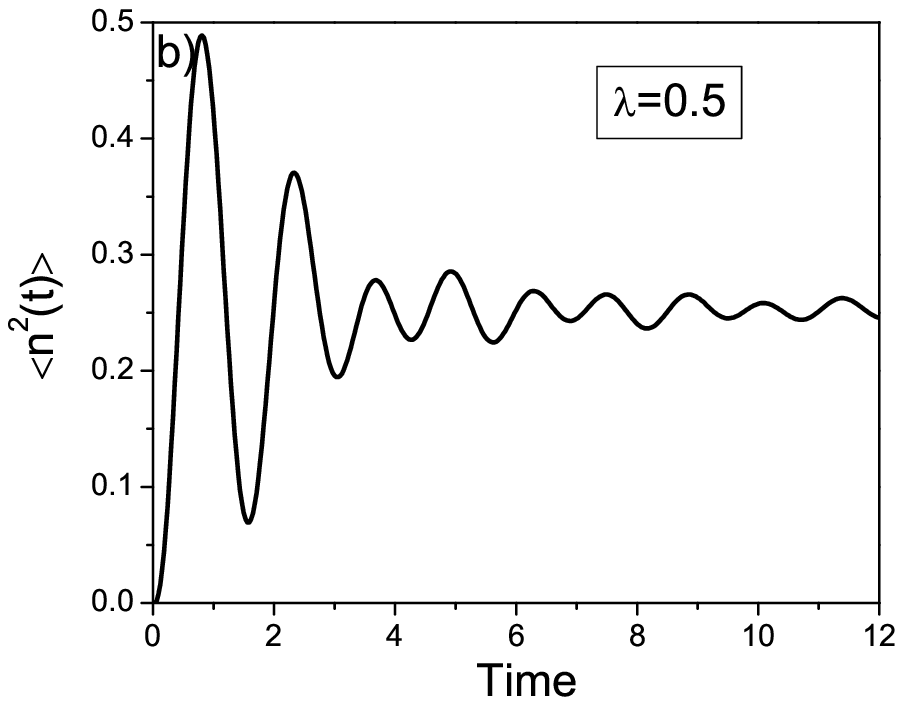}
\includegraphics[width=7.5cm]{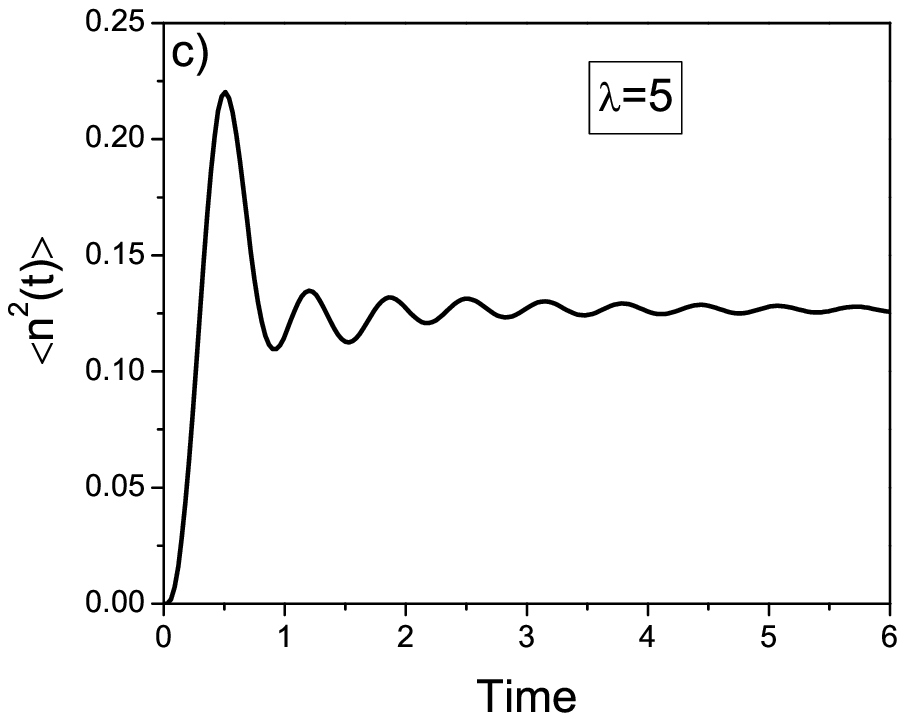}
\caption{Semiclassical variance of $n$ as a function of time. The
insert on the top graph has a different time scale.}
\label{fig1}
\end{figure}
so that the variance of $n$ approaches the steady state value of
one fourth. We note that the amplitude of oscillations decays with
time as $t^{-1/2}$ and on top of that there are beats with the
characteristic frequency $\omega_{beats}\approx 4\lambda$ (see
Fig.~\ref{fig1}). For large $\lambda$ the oscillations decay very
rapidly and $\langle n^2(t)\rangle$ quickly saturates at the
steady state value, which decreases with $\lambda$ (see
Fig.\ref{fig1}).

\subsubsection{Quantum theory}
\label{sec:quantum}

Let us now study the quantum case. The Heisenberg equations of
motion are:
\begin{equation}
{d\hat a_j\over dt}=i\,[{\cal H},\hat a_j], \label{15a}
\end{equation}
where square brackets denote commutator, $j=1,2$ and the
Hamiltonian ${\cal H}$ is given by ({\ref{1}}). It turns to be
convenient to use the following Heisenberg operators:
\begin{equation}
\left\{\begin{array}{l}
\hat\Phi=\hat a_2^\dagger \hat a_1-\hat a_1^\dagger \hat a_2,\\
\hat\Psi=\hat a_2^\dagger \hat a_1+\hat a_1^\dagger \hat a_2,\\
\hat n=\hat a_2^\dagger \hat a_2-\hat a_1^\dagger \hat a_1.
\end{array}\right.
\label{15}
\end{equation}
We introduce hats over the operators to distinguish them from
numbers appearing in the semiclassical treatment and expectation
values of the operators . It is easy to see that the following
combination \begin{equation} \hat\Psi-{\lambda\over 2 N}\hat
n^2\equiv \hat\Psi-{U\over 4 J} \hat n^2 \end{equation} commutes
with the Hamiltonian. Using this fact the system (\ref{15a}) can
be reduced to a single differential equation: \begin{equation}
{d^2 \hat n\over dt^2}+4\hat n+{2\lambda\over N}\left\{\hat
n,\hat\Psi_s\right\}_+ +{\lambda^2\over N^2}(2 \hat n^3-
\left\{\hat n, \hat n_s^2\right\}_+)=0 \label{17} \end{equation}
with the initial conditions: \begin{equation} \hat n(0)=\hat
n_s,\quad \left.{d\hat n\over dt}\right| _{t=0}=-2i\hat\Phi_s.
\label{18}
\end{equation} In the equations above $\left\{\dots\right\}_+$
denotes the anticommutator, and the subindex $s$ means
time-independent Schr\"odinger operators. We note that the second
relation in (\ref{18}) holds for all times if we use $\hat \Phi$
instead of $\hat \Phi_s$.

In the noninteracting case ($\lambda=0$) the solution of
(\ref{17}) is:
\begin{equation}
\hat n(t)=\hat n_s \cos 2t -i\hat \Phi_s \sin 2t.
\end{equation}
The initial conditions corresponding to the ground state for
$\lambda\gg \lambda_{SI}$ is $|I\rangle\equiv |N/2, N/2\rangle$.
Note that such a state is possible only if $N$ is even. The
generalization for $N$ odd is straightforward, but we will not do
it here, since our major goal is to compare quantum and
semiclassical pictures. Simple computation shows that
\begin{equation}
{n^2(t)\over N^2}\equiv {1\over N^2}\langle I|\hat
n^2(t)|I\rangle={1-\cos 4t\over 4}{N+2\over N}. \label{20}
\end{equation}
Comparing (\ref{20}) and (\ref{12}) we see that the only
difference between the semiclassical and quantum results in the
noninteracting case is the presence of an extra numerical factor
$1+2/N$ in (\ref{20}).

In the weakly interacting regime ($\lambda\ll 1$) we can neglect
terms proportional to $\lambda^2$. Then (\ref{17}) simplifies to:
\begin{equation} {d^2 \hat n\over dt^2}+4\hat n+{2\lambda\over N}\left\{\hat
n,\hat\Psi_s\right\}_+=0. \end{equation} It is very convenient to
solve this equation in the eigenbasis of $\hat\Psi_s$:
\begin{equation} |k\rangle={2^{-N/2}\over \sqrt{k! (N-k)!}}(\hat
a_{1s}^\dagger+\hat a_{2s}^\dagger )^k(\hat a_{1s}^\dagger-\hat
a_{2s}^\dagger )^{N-k}|0\rangle, \end{equation}
where $k=0,1,\dots N$. One can show that for the initial Fock
state $|I\rangle=|N/2,N/2\rangle$ the variance of $n$ is:
\begin{widetext}
\begin{equation}
{n^2(t)\over N(N+2)}={1\over 4}-{2^{2-N/2}\over
N(N+2)}\sum\limits_{k=0}^{N/2-1}
{(N\!-\!2k\!-\!1)!!\,(2k\!+\!1)!!\over (N/2-k-1)!\, k!}\cos
2t\!\left[\sqrt{1\!-\!{\lambda\over N}(4k\!+\!3\!-\!N)}\!+\!\sqrt{1\!-\!{\lambda\over
N}(4k\!+\!1\!-\!N)}\right]. \label{30}
\end{equation}
\end{widetext}

\begin{figure}
\includegraphics[width=8cm]{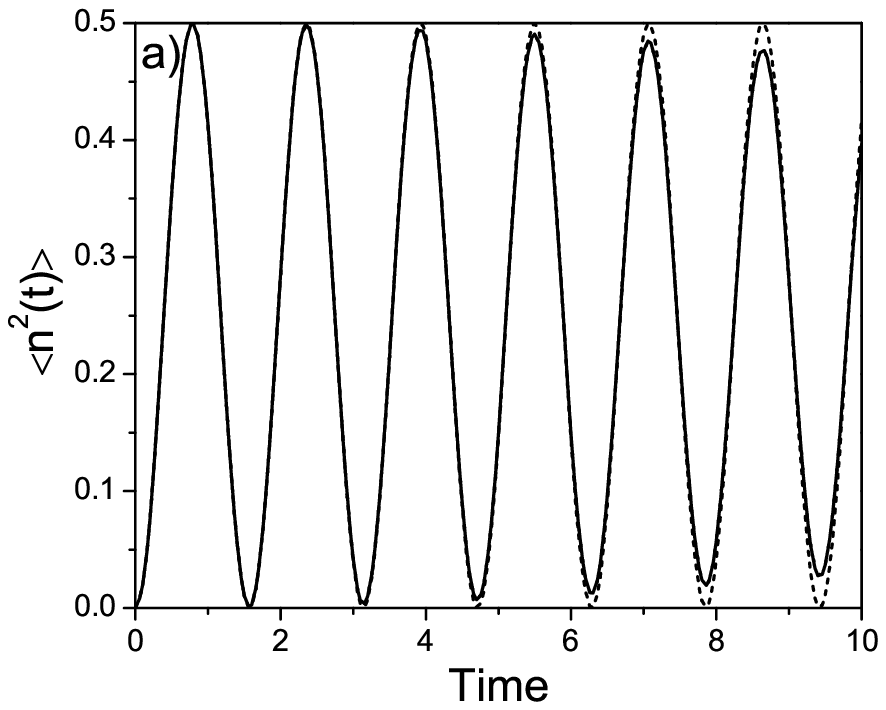}
\includegraphics[width=8cm]{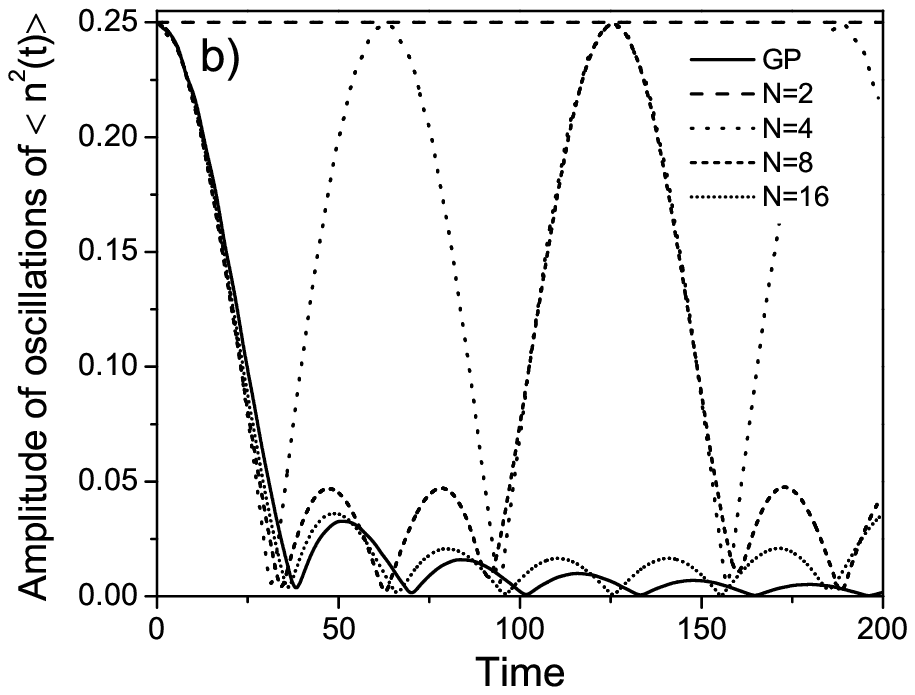}
\caption{(a) Semiclassical (solid line) and quantum variance of
$n$ as a function of time for the weak coupling case
$\lambda=0.05$. Dash line corresponds to the total number of
particles $N=2$, dot line does to $N=4$. Solid and dot line are
indistinguishable on this plot. (b) Amplitude of the oscillations
of the variance of $n$ versus time.} \label{fig2}
\end{figure}

Comparing (\ref{30}) and (\ref{13}) we see that in contrast to the
continuous integral in the semiclassical case there is a discrete
sum in the quantum. One can formally obtain (\ref{13}) from
(\ref{30}) in the limit $N\to\infty$ using Stirling's formula and
transforming the summation over $k$ to integration. It turns out
to be more convenient to normalize the variance of $n$ to $N(N+2)$
instead of $N^2$. If the total number of particles $N=2$, there is
only one term in (\ref{30}), so the oscillations are completely
undamped. For $N=4$, there are two terms and we expect perfect
beats; {\em i.e.\/} the amplitude of oscillations first goes to
zero then completely restores and so on. For $N\geq 6$ there are
several terms contributing to the sum. At relatively small time
scale $\lambda^2 t/ N\ll 1$ frequencies in different terms are
approximately equidistant: $\Delta \Omega\approx 8\lambda/N$ so
the amplitude of oscillations is a periodic function. However at a
larger time scale the phases become random and periodicity
disappears. Figure~\ref{fig2}(a) shows the comparison of the
variance of $n$ for $N=2$ and $N=4$ with the semiclassical result.
On short time scales already $N=4$ gives an excellent agreement.
In fact the semiclassical and the quantum curve (for $N=4$) are
completely indistinguishable. The behavior of the amplitude of
oscillations of $n^2$ is plotted in Fig.~\ref{fig2}(b). It is
clear that with increasing $N$, the semiclassical approximation
works for longer and longer time scales (see also
Ref.~\onlinecite{milburn}). However in a quantum system the
recurrence time is always finite, so ultimately at
$t>1/\Delta\Omega$, the semiclassical description breaks down.

\begin{figure}
\includegraphics[width=8cm]{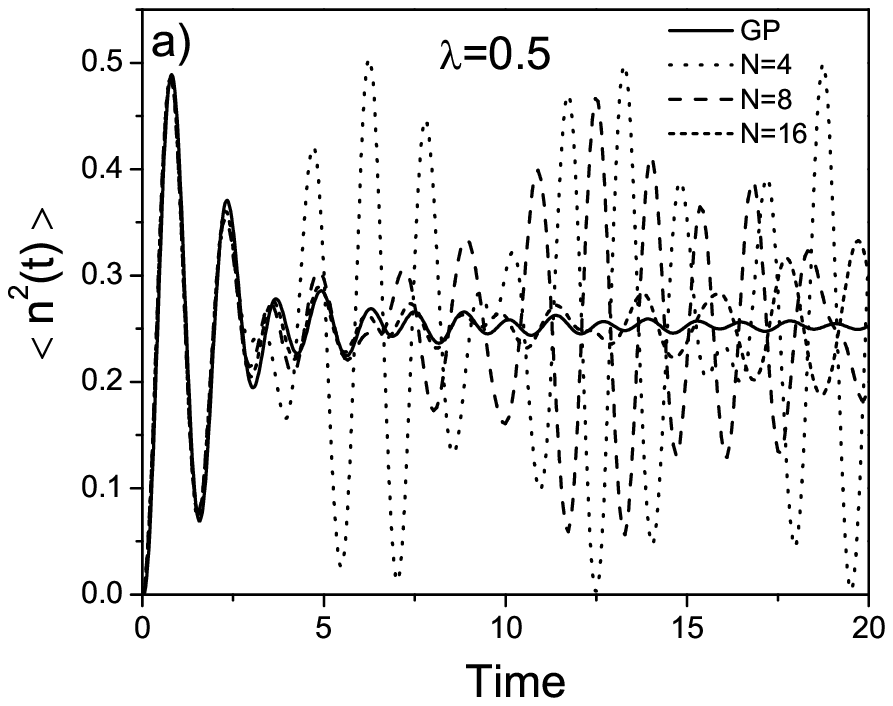}
\includegraphics[width=8cm]{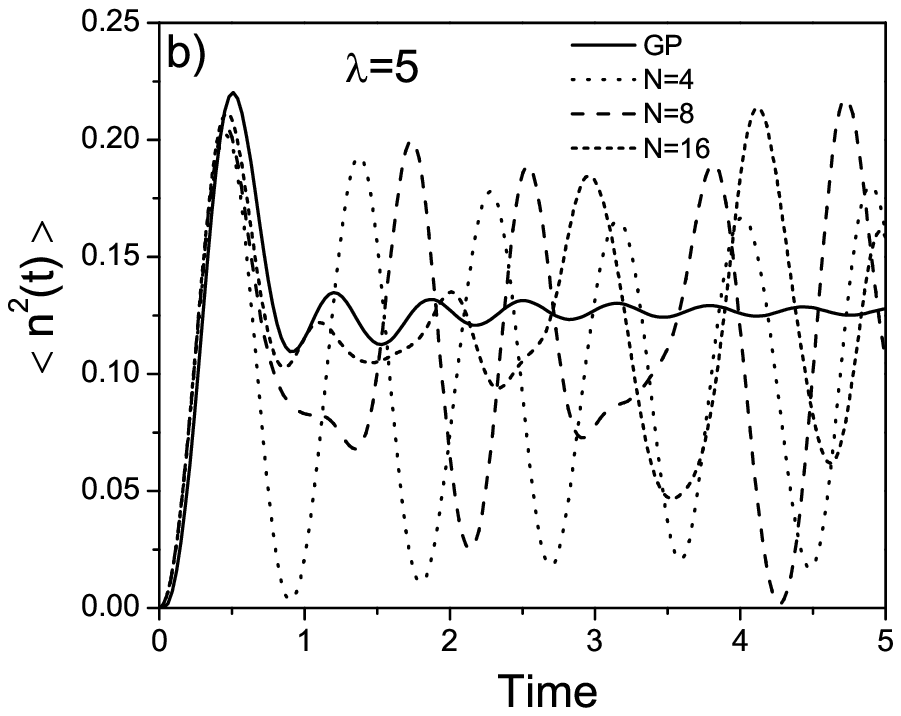}
\caption{Variance of n as a function of time for intermediate (a)
and large (b) coupling constants. Note that for larger N
semiclassical approximation works well for longer time scale, but
eventually always breaks down.} \label{fig3}
\end{figure}

In Fig.~\ref{fig3} we present the numerical solution for the case
of intermediate and strong couplings. As was discussed before for
small $N$, the amplitude of oscillations fluctuates, being
completely chaotic at large time scales. However, at sufficiently
small time, the oscillations gradually decay, approaching the
semiclassical result. At intermediate times the amplitude of the
oscillations experiences beats (compare with Fig.~\ref{fig2}).
Note that for the large coupling, the semiclassical description
breaks down very early.

\subsection{Modulated phase initial state}
\label{sec:twomod}

\begin{figure}
\includegraphics[width=7.5cm]{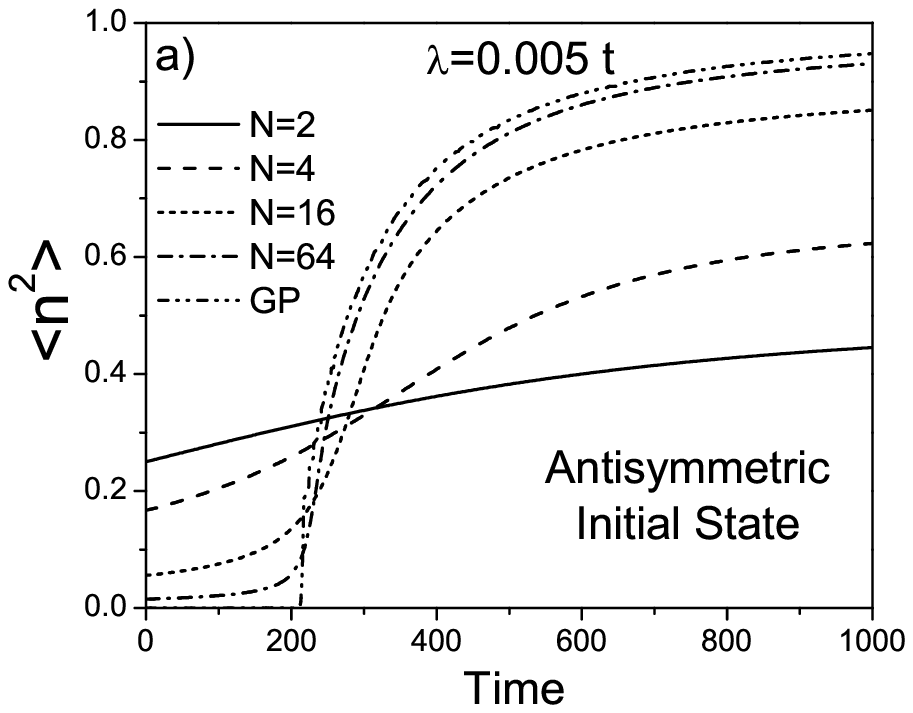}
\includegraphics[width=7.5cm]{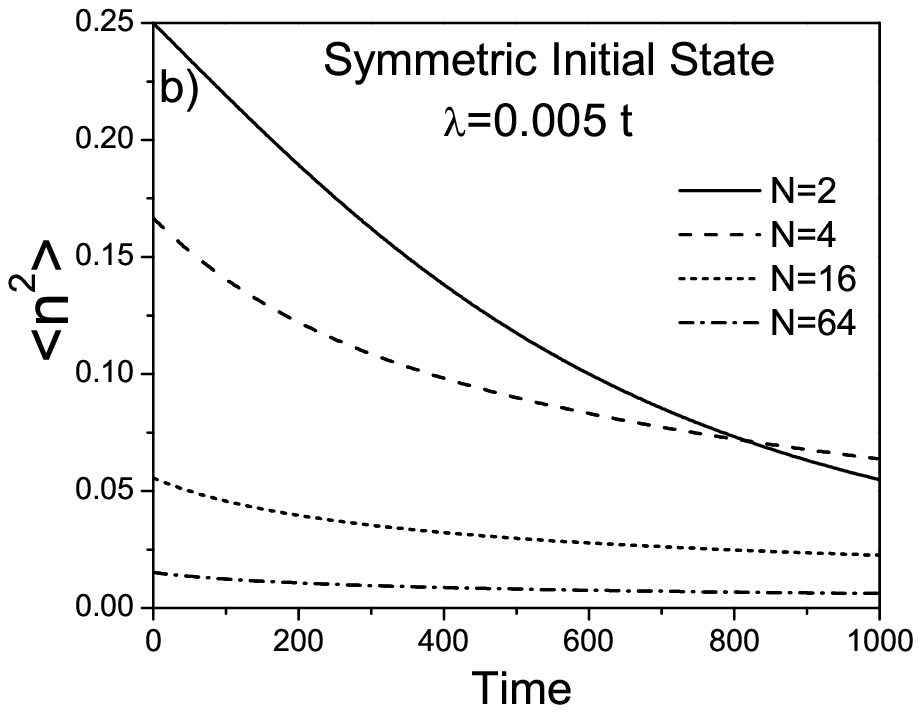}
\caption{Variance of $n$ for the two wells for adiabatically
increasing interaction $\lambda(t)$. The initial state is (a)
antisymmetric ($\phi=\pi$) and (b) symmetric ($\phi=0$).}
\label{fig4}
\end{figure}

We turn next to the initial conditions described in
Section~\ref{sec:modini}, where the initial state has a phase
order. In semiclassical picture $n$ and $\phi$ are commuting
variables and we can fix them at $t=0$ independently. For
simplicity let us consider $n_0=0$. Then from (\ref{9}) it is
obvious that only $\phi_0=0,\pi$ give the stationary solutions. As
we discussed above, $n=0$ and $\phi=0$ is automatically a ground
state for all positive values of interaction $\lambda$, therefore
it is always stable under small fluctuations. On the other hand if
$\phi_0=\pi$ then $n=0$ is (meta)stable for $\lambda\leq 1$ and
unstable for $\lambda>1$ (see Ref.~\onlinecite{Franzosi} for the
details). Suppose that we start from $\phi=\pi$, $n=0$,
$\lambda=0$ and adiabatically increase $\lambda$. Then $n^2$
remains close to zero while $\lambda$ remains smaller than
critical value. After that $n^2$ rapidly increases and the system
spontaneously goes to the Schr\"odinger cat state, where all the
bosons are either in the left or in the right well. A similar
picture holds in the quantum mechanical description. The principal
difference is that instead of a sharp transition at
$\lambda=\lambda_c$, there is a smooth crossover between the
initial and the final states. Fig.~\ref{fig4} shows the variance
of $n$ as a function of time. For comparison we consider both
symmetric ($\phi=0$) and antisymmetric ($\phi=\pi$) initial
conditions.

\section{Semiclassical Description of Multi-well Bose Gases}
\label{sec:many}

The full quantum solution of the many well case rapidly becomes
numerically prohibitive with increasing $N$, and so we will
confine our discussion in this section to the semiclassical GP
equation. From (\ref{1b}) and (\ref{poisson}) this is
\begin{eqnarray}
&&i{\partial \psi_j\over \partial
t}=-(\psi_{j+1}+\psi_{j-1})+\frac{V_j}{J}\psi_j+
\lambda|\psi_j|^2\psi_j,\label{31}
\end{eqnarray}
The equilibrium number of bosons in the central well ($j=0$)
is $N$, and so $|\psi_0|^2=1$ in the Mott insulating
ground state.

We divide our discussion according to the initial conditions
considered in Section~\ref{sec:intro}.

\subsection{Mott insulating initial state}
\label{sec:manymott}

We will compute the correlation function $\mathcal{D}_g (t)$
defined in (\ref{defDg}) for two limiting possibilities for the
weight function $g$: $g(j)=\delta_{j,1}$ and $g(j)=\,{\rm const}$,
where in the former (latter) case one computes the nearest
neighbor (global) phase correlation. Using the GP equations
(\ref{31}) we can show that \begin{eqnarray}
{d\mathcal{D}_g(t)\over dt}=&& i\sum_{j\neq \ell}\left(V_j+\lambda
|\psi_j(t)|^2\right) g(|j-\ell|)\nonumber\\
&&\times
\left(\psi_j^\star(t)\psi_\ell(t)-\psi_\ell^\star(t)\psi_j(t)\right).
\label{33}
\end{eqnarray}
Note that for uniform potential $\mathcal{D}_g (t)$ changes only due to
the interaction. In this case, the ratio ${\cal D}_g (t)/\lambda$ has
a finite limit at $\lambda\to 0$. We will consider the solution
for $\mathcal{D}_g (t)$ with and without an external
potential in the following subsections.

\subsubsection{No external potential and periodic boundary conditions}
\label{sec:pbc}

Let us assume that the lattice forms a periodic array of quantum
wells and there is no external potential ($V_j\equiv~0$). For the
nearest neighbor correlation similarly to the two well case it is
easy to show that
\begin{equation}
{\cal D}_g(t)\equiv\sum_j
\psi_j^\star\psi_{j+1}+\psi_{j+1}^\star\psi_j = {\lambda\over 2}
\sum_j (|\psi_j|^2-1)^2. \label{32}
\end{equation}

This equation shows that the nearest neighbor coherence is
proportional to the product of the coupling constant and sum of
the variances of number of bosons in each well. From the previous
section we can expect that if the interaction is weak, then
variances of $n_j$ at short time scales will be fluctuating and
governed by the noninteracting tunnelling Hamiltonian. With
increasing time the interaction will suppress the fluctuations
leading to some steady state. In the noninteracting case,
(\ref{31}) is just an ordinary Schr\"odinger equation. with
eigenstates
\begin{equation} \psi_k(j)={1\over \sqrt M}\, {\rm e}^{2\pi i k j/N},
\end{equation}
corresponding to the eigenenergies
\begin{equation} E_k=-2 \cos {2\pi k\over M}.
\end{equation}
Here $M$ is the number of wells.  Expanding the initial insulating
state in terms of the eigenstates defined above and propagating
them in time we obtain \begin{equation} \sum\limits_{j=1}^N
(|\psi_j(t)|^2-1)^2=M \left(1- \sum_j |F(j,t)|^4 \right),
\label{35} \end{equation} where \begin{equation} F(j,t)={1\over M}
\sum\limits_{k=0}^{N-1} {\rm e}^{2i\left(\pi k j/M + t \cos 2\pi
k/M\right)}. \end{equation}

For several different values of $M$ the function ${\cal
D}_g^{\,M}(t)$ at vanishing $\lambda$ is:
\begin{widetext}
\begin{eqnarray}
&&{\cal D}_g^{\,2}(t)={\lambda\over 2}\sin^2 2t,\\
&&{\cal D}_g^{\,3}(t)={8\lambda\over 9}(2+\cos 3t)\sin^2 {3\over 2}t,\\
&&{\cal D}_g^{\,4}(t)={\lambda\over 4}(7+\cos 2t)\sin^2 2t,\\
&&{\cal D}_g^{\,5}(t)={4\lambda\over 25}(10-2\cos\sqrt{5}t-\cos
\sqrt{5}t-2\cos{5\over 2} t\,\cos{3\sqrt{5}\over 2}
\cos\sqrt{5} t),\\
&&{\cal D}_g^{\,6}(t)={\lambda\over 36}(63-8\cos t-12\cos
2t-24\cos 3t-6\cos 4t-
12\cos 6t - \cos 8t),\\
&&{\cal D}_g^{\, M}(t)\to {M\lambda\over
2}\left(1-J_0(t)^4-2\sum_{m=1}^{\infty}J_m(t)^4\right)\quad\mbox{at}\quad
M\to\infty.
\end{eqnarray}
\end{widetext}

Clearly ${\cal D}_g^{\,M}(t)$ is a periodic function only for
$M=2,3,4,6$ (this is, in fact true, not only for the nearest
neighbor case). For many wells the number of harmonics
contributing to the variance of $n$ becomes large and oscillations
become more chaotic and weaker in amplitude. In the limit
$M\to\infty$, ${\cal D}_g^{\,M}(t)$ is a monotonically increasing
function. If we add the interaction, then the overall picture
remains similar to the two well case. Namely, for small $\lambda$
the amplitude of oscillations slowly decays in time. For strong
interaction, the variance of $n$ reaches steady state value in a
very short time scale.

In the opposite to nearest neighbors limit $g(|j-\ell|)={\,\rm
const}$, one can show that at $\lambda\to 0$
\begin{widetext}
\begin{equation}
{\cal D}_{g}^{\,M}(t)\to {2\lambda\over M}\sum_{k\neq m =0}^{N-1}
{\sin^2 t\left(1+\cos(2\pi k/M)-\cos(2\pi m/M)-\cos(2\pi
(k-m)/M)\right)\over 1+\cos(2\pi k/M)-\cos(2\pi m/M)-\cos(2\pi
(k-m)/M)}.\label{45}
\end{equation}
For example
\begin{eqnarray}
&&{\cal D}_{g}^{\,2}(t)={\lambda\over 2}\sin^2 2t,\\
&&{\cal D}_{g}^{\,3}(t)={\lambda\over 45}(3-2\cos 3t-\cos 6t),\\
&&{\cal D}_{g}^{\,4}(t)={\lambda\over 160}(13-12\cos 4t-\cos 8t),\\
&&{\cal D}_{g}^{\,6}(t)=\lambda{1\over 240}(33+16\cos t-24\cos
2t-8\cos 6t - \cos 8t),\\
&&{\cal D}_{g}^{\, M}(t)\to {M\lambda\over 2\pi^2}
\int_0^{2\pi}\int_0^{2\pi} d\theta_1 d\theta_2 {\sin^2t(1+\cos
\theta_1-\cos\theta_2-\cos (\theta_1-\theta_2))\over 1+\cos
\theta_1-\cos\theta_2-\cos (\theta_1-\theta_2)}
\quad\mbox{at}\quad M\to\infty.
\end{eqnarray}
\end{widetext}

\begin{figure}
\includegraphics[width=8cm]{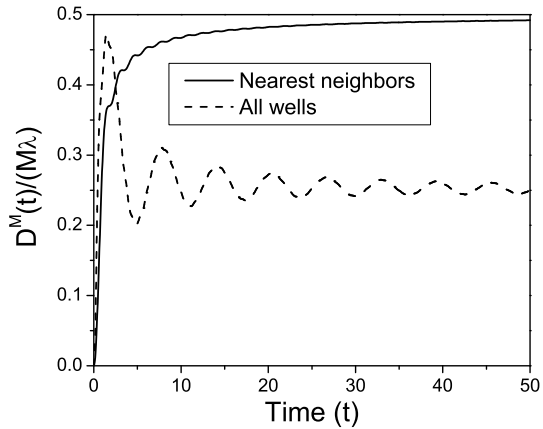}
\caption{Time dependence of the coherence ${\cal D}_g(t)$ for the
weakly interacting Bose gases at large number of wells
($M\to\infty$). Note that nearest neighbor correlation rapidly
saturates, while the global coherence exhibits oscillations.}
\label{fig5}
\end{figure}

The behavior of ${\cal D}_g(t)$ at large $M$ is very different for
nearest neighbor and global correlations (see Fig.~\ref{fig5}).
While the former rapidly reaches a steady state value, the latter
oscillates in time. Indeed the denominator in (\ref{45}) selects
only low frequency harmonics in ${\cal D}_g$, freezing out high
frequency oscillations, especially at longer time scales.
\begin{figure}
\includegraphics[width=8cm]{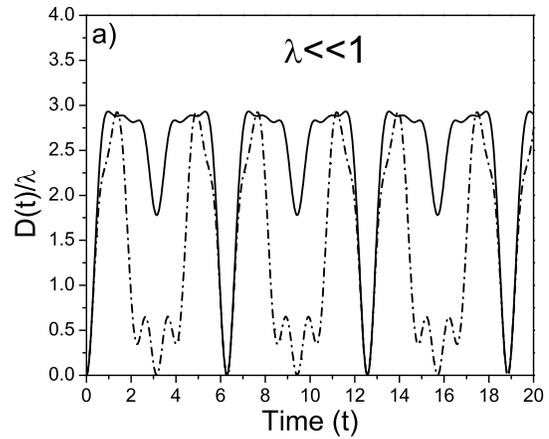}
\includegraphics[width=8cm]{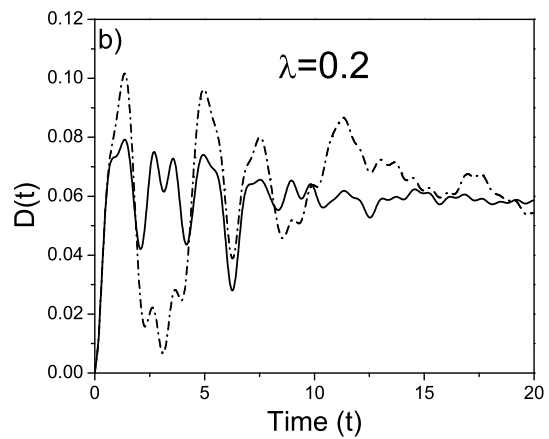}
\includegraphics[width=8cm]{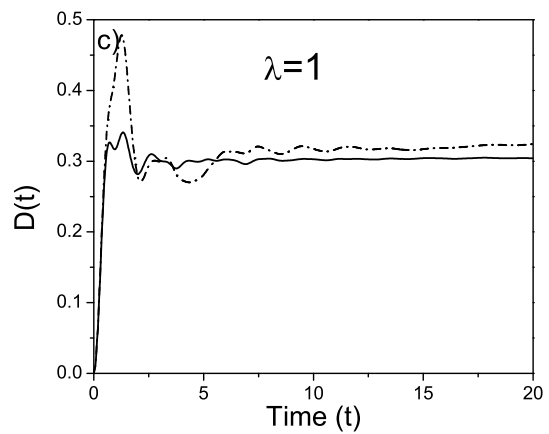}
\caption{Time dependence of ${\cal D}_g(t)$ for 6 wells; solid and
dash lines correspond to nearest neighbor and global correlations
respectively. Without interaction ($\lambda\to 0$) ${\cal
D}_g(t)$ shows regular periodic behavior in time. Nonzero
interactions leads to decay of oscillations. High frequency
oscillations of global correlation function are effectively
suppressed.} \label{fig6}
\end{figure}
\begin{figure}
\includegraphics[width=8cm]{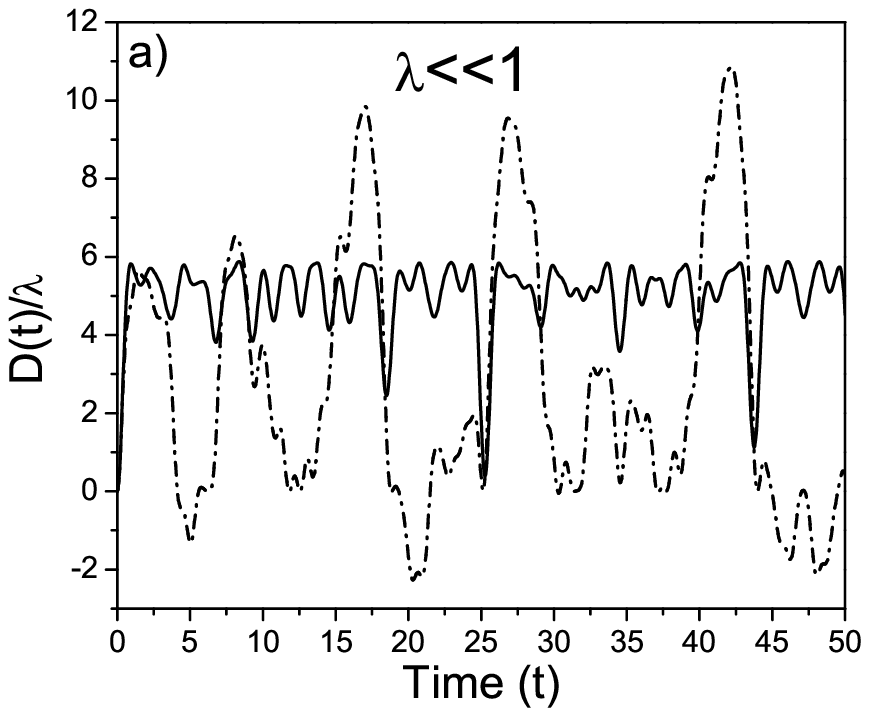}
\includegraphics[width=8cm]{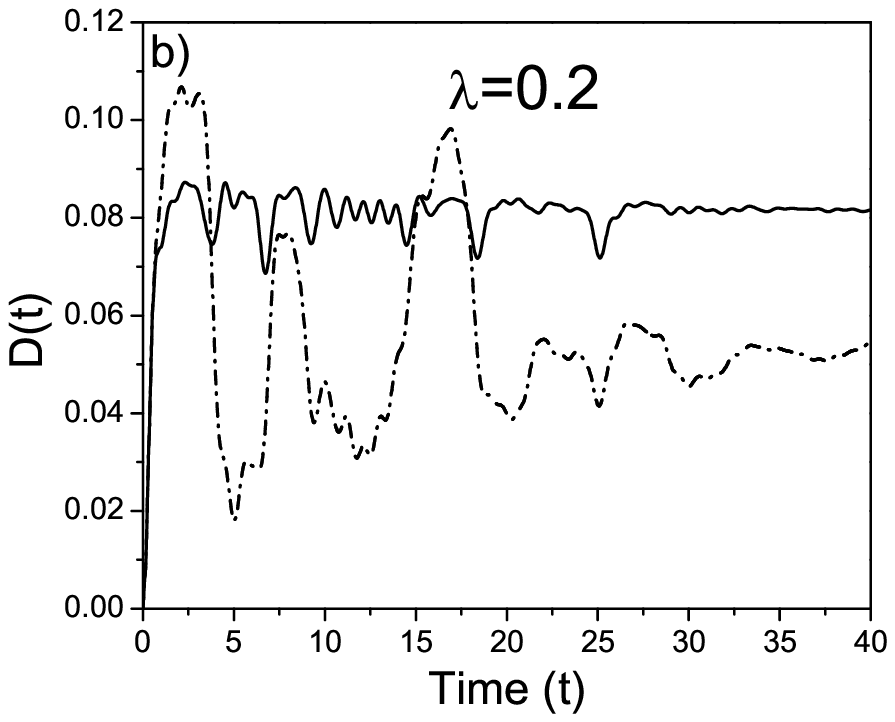}
\includegraphics[width=8cm]{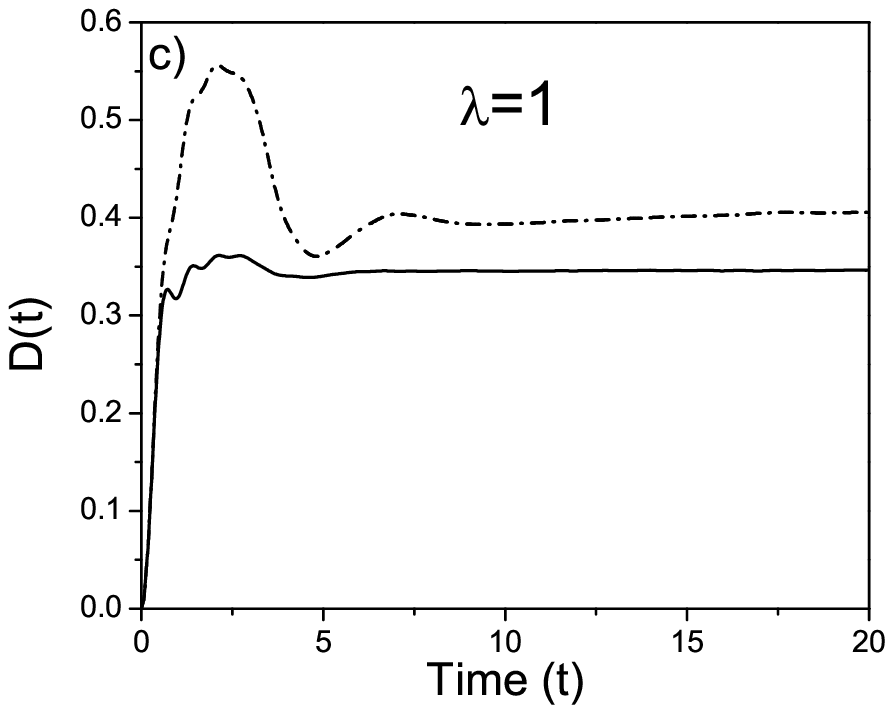}
\caption{Same as in Figure (\ref{fig6}) but for 12 wells. Without interaction
oscillations are chaotic. Low frequency dominate the global correlation
function here as well.} \label{fig6a}
\end{figure}

Figs.~\ref{fig6} and~\ref{fig6a} show ${\cal D}_g(t)$ for six and
twelve wells respectively. Six wells give periodic time
dependence, while $N=12$ corresponds to chaotic behavior. Note
that in all cases high frequency modes are suppressed for the case
of global phase correlations.

\subsubsection{Parabolic confining potential}
\label{sec:parabola}

\begin{figure}
\includegraphics[width=8cm]{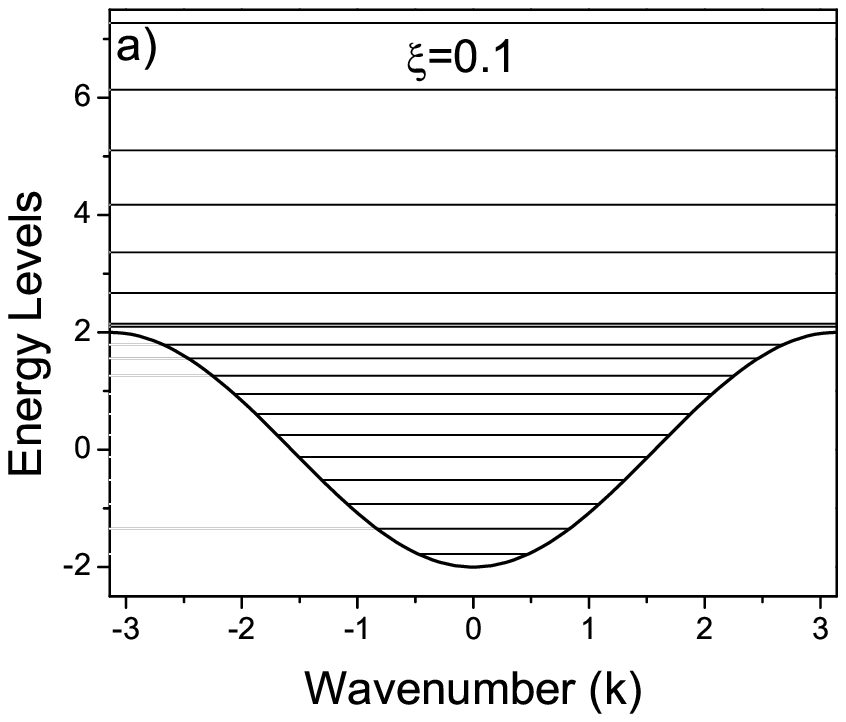}
\includegraphics[width=8cm]{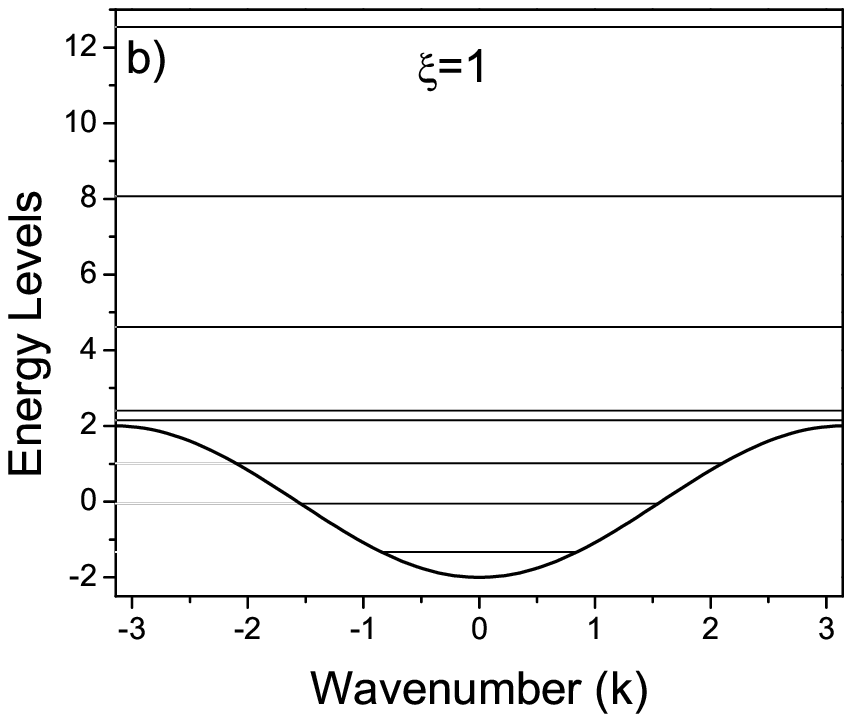}
\caption{Energy spectrum of coupled noninteracting Bose gases in a
weak (a) and intermediate (b) parabolic potential.}
\label{fig7}
\end{figure}

So far, we have considered the rather hypothetical situation of
quantum wells sitting on a ring. However, usually one achieves
confinement using a trap, which is equivalent to a nonuniform
external potential $V_j$ in (\ref{31}). The most common shape of
this potential is parabolic ($V_j\propto j^2$) and we focus on
this case, although the analysis of other potentials is similar
and straightforward. As before, we will first study the
non-interacting system: ($\lambda=0$).
\begin{equation} i{d\psi_j\over
dt}=-(\psi_{j+1}+\psi_{j-1})+{\xi j^2\over 2}\psi_j.
\end{equation}
This is a
linear Schr\"odinger equation with stationary states found from
\begin{equation}
E\psi_j=-(\psi_{j+1}+\psi_{j-1})+{\xi j^2\over 2}\psi_j.
\end{equation}
In
the Fourier space the same equation looks more familiar:
\begin{equation}
E\psi(k)=-2\cos k\,\psi(k)-{\xi\over 2}{d^2\psi(k)\over dk^2},
\end{equation}
describing the motion of an one-dimensional particle of mass
$\xi^{-1}$ living on a circle with the external potential
$U(k)=-2\cos(k)$. Note that the same type of equation describes
Josephson junctions with charging energy. If the parabolicity is
weak ($\xi\ll 1$), then the bosons form closely spaced extended
states at low energies. In the Fourier space this is equivalent
to having a heavy particle in the $-2\cos k$ potential. With a
good accuracy one can describe the energy spectrum inside such a
well using the WKB approximation. This is justified both for low
energies, where $-2\cos k\approx -2+k^2$ and the WKB gives the
exact energy spectrum and for high energies WKB works well for
any potential. In fact there is a little subtlety near energy
close to $2$, since the potential there is almost flat and can
not be approximated by a linear function, but this is not very
important. So the approximate WKB spectrum is given by
\begin{eqnarray}
&&\int_{-\pi+\cos^{-1} E/2}^{\pi-\cos^{-1} E/2}\sqrt{{2\over
\xi}(E+2\cos k)}\, d k=\pi
(n+1/2)\nonumber\\
&&\int_{-\pi}^{\pi}\sqrt{{2\over \xi}(E+2\cos k)}\, d k=2\pi n,
\end{eqnarray}
where the top (bottom) equation corresponds to $E<2$ ($E>2$). In
the first equation even or odd $n$ describes even and odd states
(in both real and reciprocal space), respectively. For energies
$E>2$, the second equation gives complete degeneracy between even
and odd energy levels. In real space roughly all states with $E>2$
are localized in individual wells, and degenerate while those with
$E<2$ are spread through many wells. Fig.~\ref{fig7}(a) briefly
summarizes this discussion showing the exact spectrum for
$\xi=0.1$ (The WKB result is indistinguishable by eye from this
graph).
Clearly the low energy levels are approximately equally spaced,
revealing the famous property of a harmonic potential, the spacing
decreases as the energy approaches $2$, and starts linearly
increasing for $E>2$ as in a usual square well. If $\xi\ge 1$ then
bosons become localized within individual wells and their energies
follow external potential. The crossover from weak to strong
parabolicity is a finite system analog of the Anderson transition.
It is important to note that this is a purely semiclassical
transition in this case, because it is derived in the
Gross-Pitaevskii picture. The ``quantum mechanics'' here
originates from the wave nature of the classical field $\psi$. If
the average number of bosons per well is much larger than one,
then the semiclassical picture, where number of bosons and their
phase commute, holds until the typical fluctuations of $\psi^2$
becomes of the order of $1/N\ll 1$. This occurs deep inside the
insulating regime, where the energy in GP approach is anyway
almost phase independent.

\begin{figure}
\includegraphics[width=8cm]{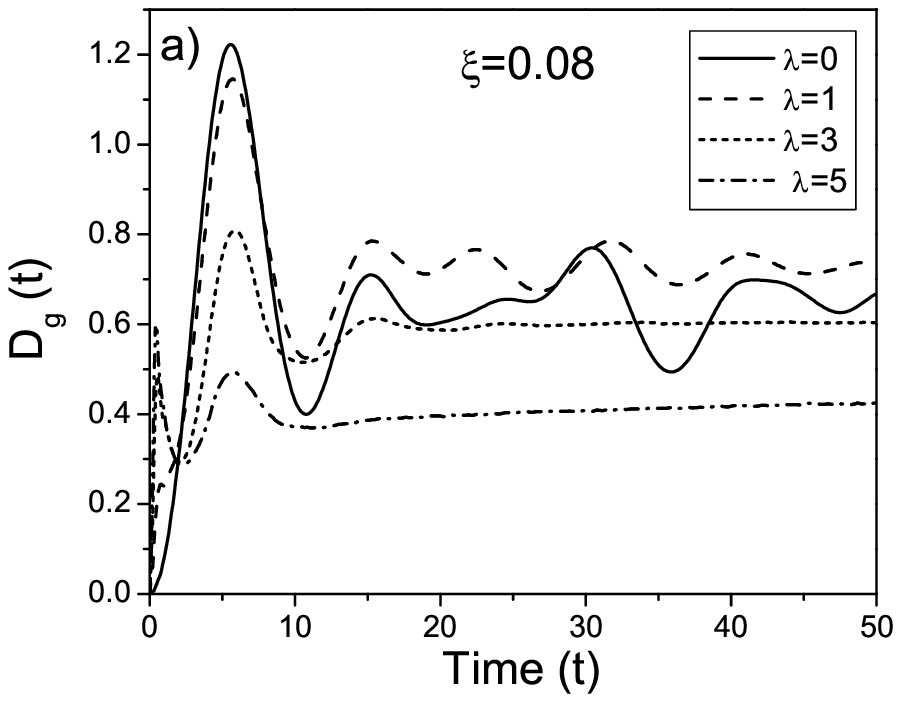}
\includegraphics[width=8cm]{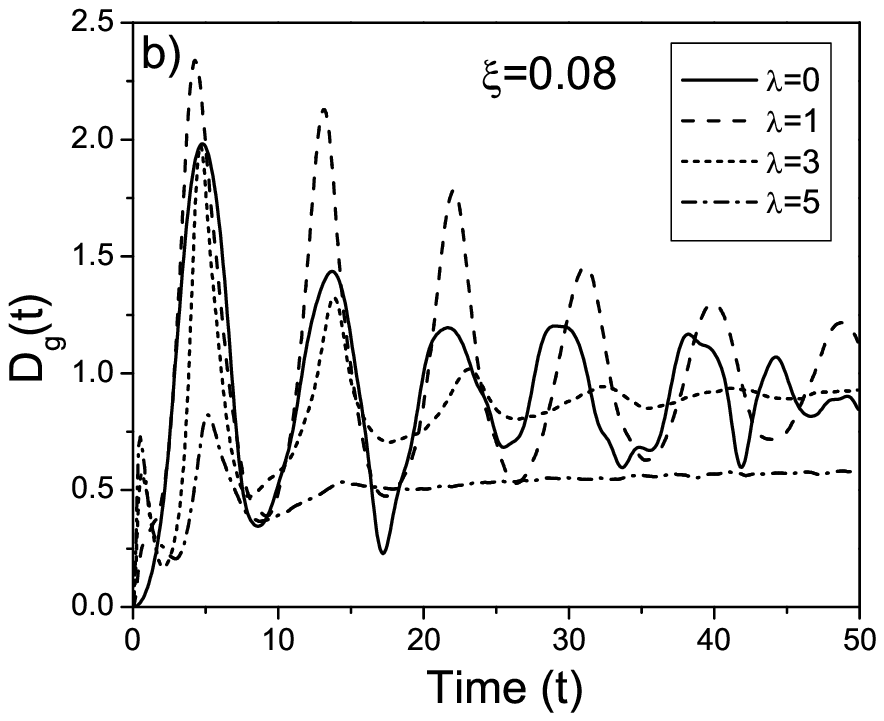}
\caption{Time dependence of ${\cal D}_g$ for the nearest neighbor
correlation (a) and global correlation (b). The period of
oscillations scales as $1/\sqrt{\xi}$ and the amplitude is finite
even without interaction $\lambda=0$. At large $\lambda$ ${\cal
D}_g(t)$ saturates very fast similarly to the flat potential. At
intermediate coupling $\lambda \sim 1$, however, the oscillations
become more pronounced than in the noninteracting regime.}
\label{fig8}
\end{figure}

After deriving the energy spectrum we can proceed with study of
the dynamics of the condensate. Note that (\ref{33}) yields that
time derivative of ${\cal D}_g(t)$ is not equal to zero even
without interaction ($\lambda=0$). Therefore we anticipate that
the results for the parabolic and flat potentials will be strongly
different, at least in the weakly interacting regime. If the
initial phases are uncorrelated then it is not hard to show that
at $\lambda=0$
\begin{widetext}
\begin{equation}
{\cal D}_g(t)=2 \sum_{j\neq \ell} V(j) g(|j-\ell|)
\sum_{p,\alpha,\beta} N_0^p \psi_\alpha^\star
(j)\psi_\alpha(p)\psi_\beta^\star(p)\psi_\beta(\ell) {\sin^2
{E_\beta-E_\alpha\over 2}t\over E_\beta-E_\alpha}, \label{56}
\end{equation}
\end{widetext}
where $N_0^p$ is the initial number of Bosons in the well number
$p$, $\psi_\alpha$ and $E_\alpha$ are the eigenfunction and energy
of the level $\alpha$ respectively . If starting from the ground
insulating state then
\begin{eqnarray}
N_0^p=1-{V_p\over \mu}\quad\mbox{for}\quad V_p<\mu,\\
N_0^p=0 \quad\mbox{for}\quad V_p>\mu, \end{eqnarray} with $\mu$
being a chemical potential. Let us make few comments about
({\ref{56}}). Levels $\beta$ and $\alpha$ must have the same
parity, meaning the lowest harmonic contributing to the sum will
be $\omega_{\rm
min}=2\,\mbox{min}_\alpha\,(E_{\alpha+2}-E_\alpha)>0$. Because
$N_0^p$ is centered near the bottom of the well, only levels with
delocalized wavefunctions will contribute to the sum. In
particular, degenerate levels with $E>2$ can be safely thrown
away. If $g(|j-\ell|)$ is constant, then summation over $m$
ensures that the major contribution comes from $\beta=0$;
therefore ${\cal D}_g(t)$ contains mostly harmonics with
$\omega=E_2-E_0$, $\omega=E_4-E_0$, etc., with the strongest
weight at the smallest frequency. Note that at small energies and
weak parabolicity the lowest energy levels are approximately
equally spaced, therefore the whole expression for ${\cal D}_g(t)$
will be a quasi-periodic function of a frequency $\omega\approx
E_2-E_0$. However, because this equidistance is not exact, the
periodicity will be only approximate, and at a short time scale
the amplitude of oscillations will slowly decay. On the contrary
for the nearest neighbor phase coherence
$g(|j-\ell|)=\delta_{j,\ell\pm 1}$ neither $\beta$ nor $\alpha$
are bounded to the ground state and we expect that all kinds of
allowed frequencies $E_\alpha-E_\beta$ will give contributions.
Clearly in this case dephasing occurs much earlier and the
amplitude of oscillations is much weaker. Also the characteristic
frequency of the oscillations for the nearest neighbor case will
be somewhat larger than that for the global case since the level
separation decreases with energy. Fig.~\ref{fig8} shows time
dependence of ${\cal D}_g$ for nearest neighbor and global
correlations at the parabolicity $\xi=0.08$. From the above
analysis we should expect the major oscillations at the period
\begin{equation} T= {2\pi\over E_2-E_0}\approx {\pi\over
\sqrt{2\xi}}\approx 8, \label{paraperiod} \end{equation} which is
indeed very close to the numerical value.

Interesting things happen if we turn on the interaction. In
particular, if $\lambda$ is of the order of one, the oscillations
become much more pronounced and smooth compared to noninteracting
case (see Fig.~\ref{fig8}). This is at first quite an unexpected
result, since we know that the interaction leads to decoherence
and saturation of ${\cal D}_g$. However this is not the whole
story. In the previous analysis we saw that at least for the
${\cal D}_g(t)$, interaction ``kills'' high frequency
contributions first. But that is precisely what we need for
harmonic behavior. So crudely speaking, small or intermediate
interaction removes harmonics causing dephasing of the
noninteracting function ${\cal D}_g$. If interactions become
strong $\lambda\gg 1$, then the noninteracting picture is
irrelevant and we come back to the usual behavior with fast
saturation of ${\cal D}_g$. Notice from Fig~\ref{fig8}, that the
noninteracting and interacting pictures are quite different at
small time. This can be also understood naturally as a result of
interplay of many harmonics at early stage of the evolution. Hence
we expect that the typical time scale for the first maximum in the
interacting problem will be of the order of the tunnelling time,
which is much shorter than inverse level spacing. However at later
times only slow harmonics survive leading to slight modifications
of the noninteracting picture.

\subsection{Modulated phase initial state}
\label{sec:manymod}

\begin{figure}
\includegraphics[width=8cm]{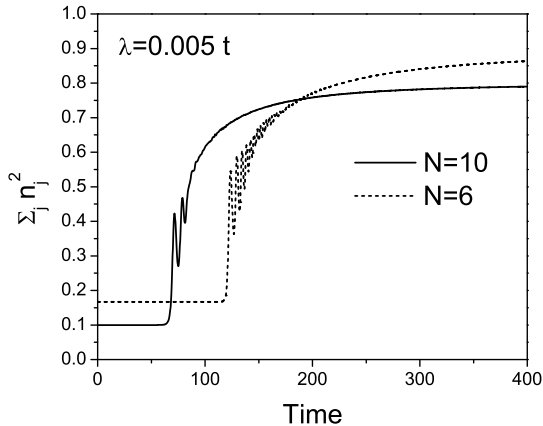}
\caption{Sum of the squares of number of bosons in different
lattice cites (with normalization $\sum_j n_j=1$). Clearly uniform
distribution is stable until interaction is smaller than the
critical value~\ref{328}. At $t\to\infty$ we have
$\sum_j~n_j^2~\to~1$ implying that all the bosons populate one of
the wells.} \label{fig9}
\end{figure}

It is also straightforward to generalize the discussion of
Section~\ref{sec:twomod} to the case of the periodic lattice.
Namely, if the number of wells is even, then the state with a
relative phase shift $\pi$, and equal numbers of bosons in the
wells, is metastable for weak interaction. If $\lambda$ increases
gradually, then when it reaches a critical value $\lambda_c$, this
state becomes unstable~\cite{Wu,Smerzi}. The critical value of
$\lambda$ can be found from the linear analysis of (\ref{31})
near the $\pi$ state~\cite{Wu,Smerzi}:
\begin{equation}
\psi_j(t)\approx{\rm e}^{i\pi j-i (2+\lambda) t}\left(1+u{\rm
e}^{iqj-i\omega t}+v^\star {\rm e}^{-iqj-i\omega t}\right),
\label{328a}
\end{equation}
where $u$ and $v$ are the small amplitudes and $q\neq 0$ is the
wave vector of the perturbation. Substitution of this expansion
into (\ref{31}) gives the following secular equation for the
eigenfrequencies $\omega$:
\begin{equation}
\left|\begin{array}{cc} \omega+2-2\cos q -\lambda & -\lambda\\
-\lambda & -\omega+2-2\cos q-\lambda
\end{array}\right|=0,
\end{equation}
which has two solutions
\begin{equation}
\omega=\pm 2\sqrt{(1-\cos{q})^2-\lambda (1-\cos q)}
\end{equation}
Clearly $\omega$ is real if $\lambda<1-\cos q$. Otherwise,
fluctuations with wavevector $q$ become unstable since the
frequency becomes complex. The lowest nonzero $q$ for the periodic
boundary conditions is $2\pi/M$, so the critical value of the
interaction, where the $\pi$ state becomes the saddle point
rather than local minimum is
\begin{equation}
\lambda_c=2 \sin^2 {\pi\over M}.
\label{328}
\end{equation}
Similar to
the two well case, the bosons undergo a spontaneous transition to
the superposition of states, where all of them are in one of the
wells. The time dependence of the variance of $N$ is analogous to
that plotted on the top graph of Fig~\ref{fig4} (see
Fig~\ref{fig9}). We remark that a ``slow'' or adiabatic increase
of interaction must be understood carefully. In the GP picture, an
adiabatic increase of interaction means that the characteristic
time scale is much smaller than the tunnelling time: $(d\ln
\lambda/dt\ll 1)$. On the other hand, for the quantum problem
adiabaticity would imply that $d\ln\lambda/dt$ is much smaller
than the level spacing, which is proportional to inverse number of
bosons. If the interaction is increased adiabatically in the
quantum mechanical sense, then the system would follow the local
minimum of the metastable state, and when $\lambda$ becomes larger
than the critical value, it will undergo a spontaneous transition
to the dipole state (or a superposition of the dipole states) with
broken translational symmetry.

\section{Conclusions}
\label{sec:conc}

We have studied the non-equilibrium temporal behavior of coupled
bosons in a lattice. We predicted dynamical restoration of the
phase coherence after a sudden increase of the tunnelling in a
system initially in a Mott insulating state. In the strongly
interacting case, $\lambda\gg 1$, the coherence reaches a steady
state rapidly (within a Josephson time). On the other hand, time
evolution in the weakly interacting regime $\lambda\lesssim 1$
depends strongly on the details of the confining potential. We
predicted that in a parabolic potential $V_j=\xi j^2/2$ the
coherence exerts decaying oscillations with period $T\propto
1/\sqrt{\xi}$ (see (\ref{paraperiod})). The period and the
amplitude of oscillations only depend weakly on interaction in
this case. On the other hand, if the confining potential is flat,
then the oscillations are either periodic (for a particular number
of wells in a lattice) or chaotic. Here the interaction leads to
the decay of the oscillations with time. In both cases the system
ultimately reaches steady state with nonzero coherence (dynamical
Bose Einstein condensate).

For the two well case we explicitly tested the validity of GP
approach. It was shown that the mapping of the deterministic
quantum mechanical motion to the stochastic GP equations is
essentially exact for time less than the characteristic inverse
level spacing $t<N/\lambda$. Apart from the slight renormalization
of the overall constant, the mapping is excellent in this time
domain already for two bosons per well. For stronger interactions,
the semiclassical and quantum mechanical trajectories start to
depart faster, as expected.

We also considered the dynamical appearance of  ``Schr\"odinger
cat'' state under a slow increase of interaction from an initial
phase modulated $\pi$ state. The $\pi$ state is stable while
interaction is weak and becomes unstable when $\lambda>\lambda_c$.
In the GP picture, this instability leads to the symmetry
breaking, so that all the bosons spontaneously populate one of the
wells. Quantum mechanically this means that the final
configuration is the superposition of states in which bosons
occupy different lattice sites. This approach can be used
experimentally for the creation of strongly entangled states.

\begin{acknowledgments}
We are indebted to M.~Kasevich and A.~Tuchman for sharing the
results of their ongoing experiments and for numerous very useful
discussions. We thank E.~Altman, and A.~Auerbach for communicating
their results prior to publication and for an illuminating
correspondence. This research was supported by NSF Grants DMR
0098226 and DMR 0196503.
\end{acknowledgments}

\appendix
\section{Mean field ground state of the boson lattice system in a
parabolic potential} \label{app:mft}

The problem of the Mott insulator transitions for infinite arrays
of bosons have been extensively studied during the last decade,
see for example~\cite{Fisher,Sachdev,Monien}. It was
shown~\cite{Sachdev} that the mean filed calculations
qualitatively captures the two possible phases and gives a good
estimate for the phase boundary. Recently, using quantum
Monte-Carlo methods, an exact ground state for the system of
bosons in a parabolic potential was found~\cite{Troyer}. It was
shown that near the expected transition, the global
compressibility does not vanish due to the spatial inhomogeneity.
However, still the bosons form local insulating domains separated
by narrow superfluid regions. The Monte Carlo approach, though
very powerful, is incapable to solving the problem with many
bosons per well. Therefore we think that for qualitative
understanding of the ground state as a function of the interaction
strength, it is worthwhile to do a mean field calculation.

The details of the derivation of the mean field equations can be
found in Ref.~\onlinecite{Sachdev}. Here we will only outline the
principal steps.

The mean field version of the free energy, corresponding to
(\ref{1}) is
\begin{eqnarray}
{\cal H}_{mf}=&&-\sum_j J (b_j a_j^\dagger+ b_j^\star
a_j)+(V_j-\mu) a_j^\dagger a_j\nonumber\\
&&+{U\over 2} a_j^\dagger a_j(a_j^\dagger a_j-1), \label{a1}
\end{eqnarray}
where $\mu$ is the chemical potential. The variational parameter
$b_j$, corresponding to the ground state is:
\begin{equation}
b_j={\langle a_{j+1}+a_{j-1}\rangle\over 2}, \label{a2}
\end{equation}
where the average is taken in the ground state of (\ref{a1}). We
can define the order parameter
\begin{equation}
\rho={\sum_j b_j^\star b_j}.
\end{equation}

\begin{figure}
\includegraphics[width=8.44cm]{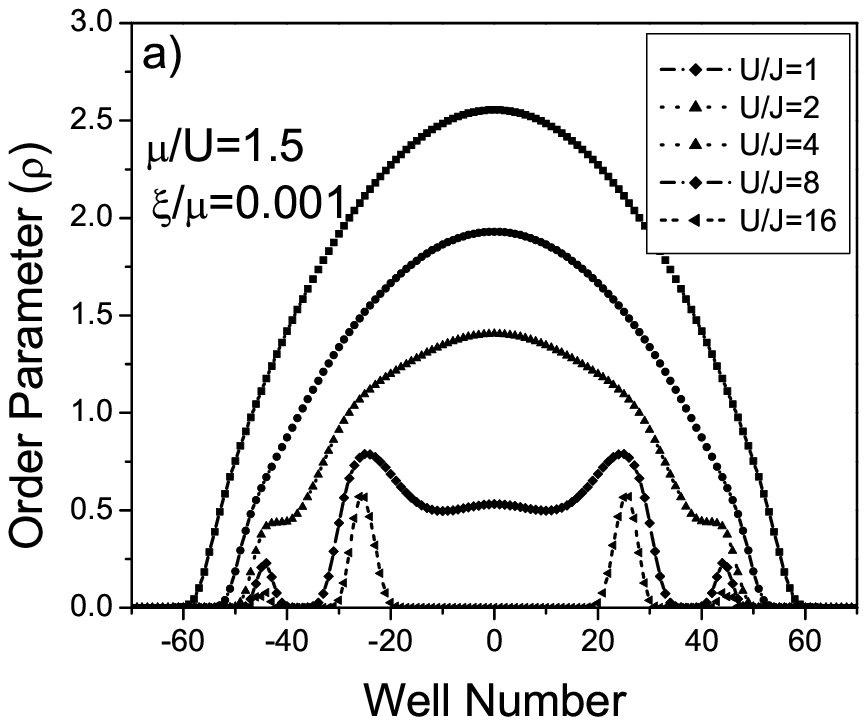}
\includegraphics[width=8.44cm]{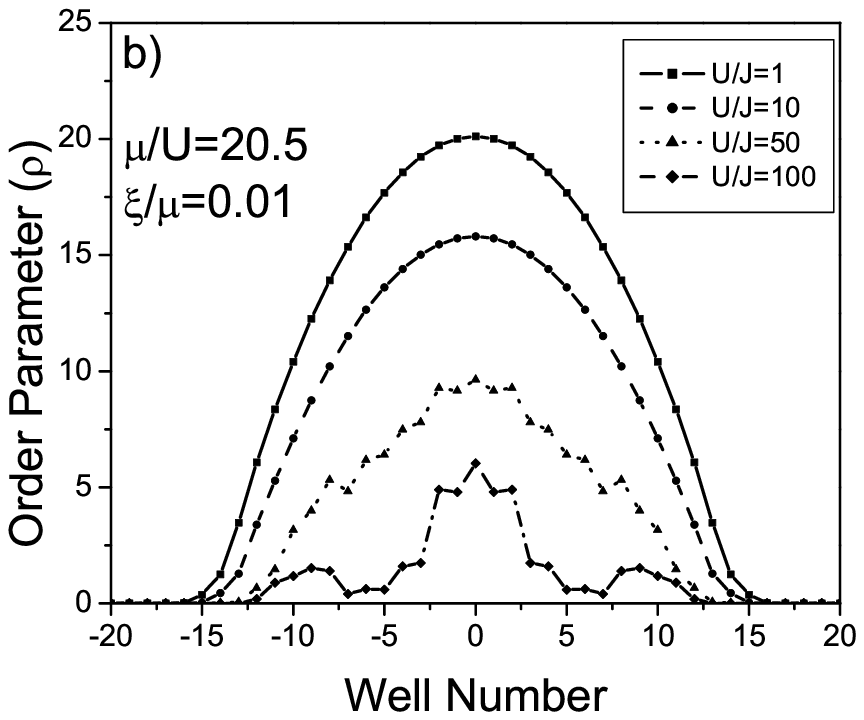}
\includegraphics[width=8.44cm]{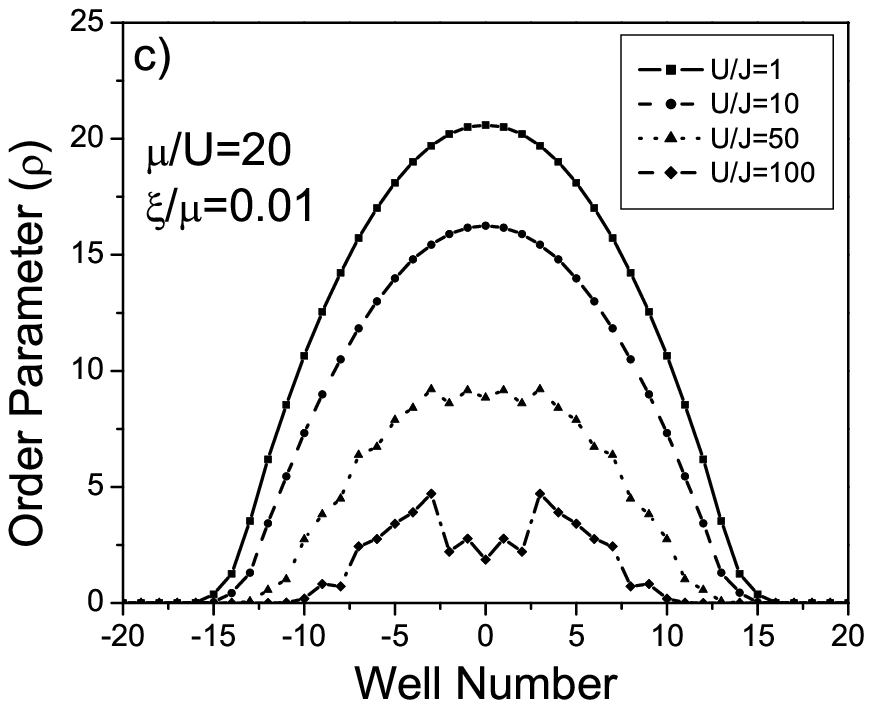}
\caption{Mean field order parameter for different interactions in
a parabolic potential ($V_j=\xi j^2/2$). Graph (a) corresponds to
few bosons per site and the other two graphs do to many bosons.}
\label{fig0}
\end{figure}

The self consistent evaluation of the mean field $b_j$ is
straightforward and the resulting order parameter is plotted in
Fig~\ref{fig0}. The graph (a) corresponds to few bosons per
lattice site. If the interaction ($U$) is strong enough, then the
order parameter forms a domain structure similar to that predicted
in~\cite{Troyer}. For a large number of bosons per well, the
quantum fluctuations start playing a role when $U$ becomes of the
order of the number of bosons in the central well ($N\approx \mu/
U$), and the smooth GP shape of boson density ($\rho$) breaks
down. For very strong interaction, the actual profile of $\rho$
becomes sensitive to small variations of the mean density of
bosons per central well.

\section{Amplitude fluctuations near the superfluid-insulator
transition} \label{app:amp}

This appendix reviews results on the damping of the amplitude
oscillation mode near the superfluid-insulator transition,
motivated by the recent paper of Altman and Auerbach \cite{aa}. As
we discussed in Section~\ref{sec:mottini}, we have considered a
system deep in the Mott insulating phase (with $\lambda \gg
\lambda_{SI}$) taken suddenly to parameters for which the ground
state was deep in the superfluid phase (with $\lambda \ll
\lambda_{SI}$), while Altman and Auerbach consider the case when
both the initial and final values of $\lambda$ were not too far
from $\lambda_{SI}$, but remained on opposite sides of it.

A key ingredient in the dynamics of the amplitude mode for
$\lambda < \lambda_{SI}$ is the damping induced by emission of the
Goldstone ``spin wave'' or ``phonon'' modes. This problem was
considered in Refs.~\onlinecite{ssrelax,sscrit}, and it was found
that the amplitude oscillations were overdamped in the $\lambda <
\lambda_{SI}$ scaling limit associated with the second-order
superfluid-insulator transition. We will review these results
below, and display expressions which also allow us to move beyond
the scaling limit to values of $\lambda$ much smaller than
$\lambda_{SI}$ (see \ref{epar2}); the amplitude mode can become
oscillatory in the latter regime \cite{ssrelax,aa}. This also
consistent with the considerations of the present paper, where we
have found that the oscillations of the superfluid coherence were
present in the parabolic multiwell case for $\lambda = 5 $ in
Fig~\ref{fig8}, but were fully overdamped for $\lambda =10$ (not
shown). We found similar behavior in the complete quantum solution
for the two-well problem---however in the latter case, the
oscillations  reappeared at very large $\lambda \sim N^2$: these
are the ``number'' oscillations of the Mott insulator, and were
also found in Ref.~\onlinecite{ssrelax}. The fate of these very
small and very large $\lambda$ oscillations in the multiwell case
near $\lambda_{SI}$ requires a treatment of the interacting
quantum dynamics: this was done in
Refs.~\onlinecite{ssrelax,sscrit}, and the results are reviewed
here.

As is well known, we can describe the superfluid-insulator
transition by the $N=2$ case of the $N$-component $\varphi^4$
field theory, where the superfluid order parameter $\psi$ in
Section~\ref{sec:intro} \begin{equation} \psi \sim \varphi_1 + i
\varphi_2
\end{equation}. The action for $\lambda$ close to $\lambda_{SI}$
is \begin{eqnarray} \mathcal{S} &=& \int d^d x d \tau \Biggl[
\frac{1}{2} \left( \nabla_x \varphi_{\alpha} \right)^2 +
\frac{1}{2 c^2} \left(
\partial_{\tau} \varphi_{\alpha} \right)^2 - \frac{(r_c + s)}{2}
\varphi_{\alpha}^2 \nonumber \\ &~&~~~~~~~~~~~~+ \frac{u}{2N}
\left( \varphi_{\alpha}^2 \right)^2 \Biggr], \end{eqnarray} where
$\alpha = 1 \ldots N$, $c$ is a velocity, $d$ is the spatial
dimensionality and $u$ is a quartic non-linearity. The coefficient
of $\varphi_{\alpha}^2$ is used to tune the system across the
transition, and the value of $r_c$ is chosen to that the
transition occurs at $s=0$ {\em i.e.} $s \sim \lambda -
\lambda_{SI}$. We assume that in the superfluid phase $\langle
\varphi_{\alpha} \rangle = N_0 \delta_{\alpha,1}$. The
oscillations of the spin-wave modes are given by the transverse
susceptibility $\chi_{\perp} (k, \omega)$, while those of the
amplitude mode are given by the longitudinal susceptibility
$\chi_{\parallel} (k, \omega)$; here $k$ is a wavevector, $\omega$
is a frequency, and the susceptibilities are defined by
\begin{eqnarray} \chi_{\perp} (k, \omega) &=& \left \langle \left|
\varphi_{2} (k, \omega) \right|^2 \right\rangle \nonumber \\
\chi_{\parallel} (k, \omega) &=& \left \langle \left| \varphi_{1}
(k, \omega) \right|^2 \right\rangle - N_0^2 (2 \pi)^{d+1}
\delta(k) \delta(\omega)~~ \end{eqnarray}

Expressions for $\chi_{\perp,\parallel}$ were given in
Refs.~\onlinecite{ssrelax,sscrit} using both perturbation theory
in $u$ and the large $N$ expansion. Here, we collect them with a
common notation, and interpret them in the present context. To
first order in $u$, the position of the critical point is
determined by \begin{equation} r_c = \frac{2u(N+2)c}{N} \int
\frac{d^{d+1} p}{(2 \pi)^{d+1}} \frac{1}{p^2} \label{rcrit}
\end{equation} where $p = (k, -i \omega/c)$ is the
$(d+1)$-dimensional Euclidean momentum. In the limit of large $N$,
but $u$ arbitrary, the value of $r_c$ is given simply by the $N
\rightarrow \infty$ limit of (\ref{rcrit}). To first order in $u$,
we obtain for $\chi_{\perp}$ \begin{equation} \chi_{\perp}^{-1}
(p) = p^2 - \frac{8csu}{N} \int \frac{d^{d+1} q}{(2 \pi)^{d+1}}
\frac{1}{q^2 + 2s} \left( \frac{1}{(p+q)^2} - \frac{1}{q^2}
\right), \label{swdamp}
\end{equation} where $q$ is also a $(d+1)$-dimensional Euclidean
momentum; at $N=\infty$ we have simply $\chi_{\perp}^{-1} (p) =
p^2$. The expression (\ref{swdamp}) describes the spin-wave
oscillations, along with their essentially negligible damping from
their coupling to the amplitude mode (as can be verified by taking
the imaginary part of the loop integral in (\ref{swdamp}) after
analytic continuation to real frequencies).

The damping in the longitudinal modes is much more severe, and we
will consider it explicitly. To first, order in $u$, we obtain the
expression \begin{equation} \chi_{\parallel}^{-1} (p) = p^2 + 2s -
\frac{4csu(N-1)}{N} \Pi (p) + \delta \chi_{\parallel}^{-1} (p).
\label{epar1} \end{equation} Here the strong damping term has been
included in $\Pi (p)$ whose explicit form is discussed below in
(\ref{Pip}), while $\delta \chi_{\parallel}$ contains additional
non-singular terms we can safely neglect. For completeness, we
give the expression for the latter \begin{eqnarray}
\delta\chi_{\parallel}^{-1} (p) &=& \frac{12uc}{N}\int
\frac{d^{d+1} q}{(2 \pi)^{d+1}} \left( \frac{1}{q^2} -
\frac{1}{q^2+2s} \right)  \\ &-& \frac{36 cs u}{N} \int
\frac{d^{d+1} q}{(2 \pi)^{d+1}} \frac{1}{(q^2 + 2s)((p+q)^2 +
2s)}; \nonumber \end{eqnarray} note that these terms always
involve coupling to an amplitude mode fluctuation (with ``mass''
$2s$) and this is the reason their contribution is non-singular.
We find below in (\ref{Pip}) that the $\Pi (p)$ contribution in
(\ref{epar1}) involves only spin-wave fluctuations and hence it
becomes very large at low frequencies, where the perturbative
expansion in (\ref{epar1}) can no longer be trusted. Fortunately,
a resummation of these singular corrections is provided by the
large $N$ expansion, which yields \begin{equation}
\chi_{\parallel}^{-1} (p) = p^2 + \frac{2s}{1 + 2 cu \Pi (p)};
\label{epar2} \end{equation} it is satisfying to check that
(\ref{epar2}) and (\ref{epar1}) are entirely consistent with each
other in their overlapping limits of validity of small $u$ and
large $N$. The expression (\ref{epar2}) was given earlier
\cite{ssrelax} in the scaling limit, which corresponds to ignoring
the 1 in the denominator because $\Pi (p)$ becomes large. The
utility of (\ref{epar2}) is that it does not have divergent
behavior at small $p$.

We turn, finally, to the expression for $\Pi (p)$, which is
\begin{eqnarray} \Pi (p) &=& \int \frac{d^{d+1} q}{(2 \pi)^{d+1}}
\frac{1}{q^2
(p+q)^2} \nonumber \\
&=& F_d |p|^{d-3}, \label{Pip} \end{eqnarray} where $F_d$ is a
numerical prefactor which is not difficult to obtain explicitly.
Notice that $\Pi (p)$ is singular as $p \rightarrow 0$ in $d<3$,
and this is the reason for the strong damping of the amplitude
mode. After analytic continuation to real frequences, we have in
$d=2$
\begin{equation} \Pi (k, \omega) = \frac{1}{8 \sqrt{k^2 -
(\omega/c)^2}}~~~;~~~d=2; \label{dampd2} \end{equation} this has a
non-zero imaginary part for $\omega
> ck$ which leads to the damping of the amplitude mode. The
expression for $\Pi (p)$ is infrared divergent in $d=1$, and this
is the signal that there is no true long-range order;
nevertheless, its imaginary part remains well defined as $d
\searrow 1$, and we find \begin{equation} \mbox{Im} \Pi (k,
\omega) = \frac{1}{4 ((\omega/c)^2 - k^2)} \theta(\omega -
ck)~~~;~~~d=1 \label{dampd1} \end{equation} which again predicts
strong damping at low frequencies. The expressions
(\ref{epar2}-\ref{dampd1}) can be used to describe the evolution
of the weakly damped amplitude mode at $\omega = \sqrt{c^2 k^2 +
2s}$ at large $s$ deep in the superfluid, to the overdamped mode
with no sharp resonance at this frequency for small $s$.

\end{document}